\documentclass[12pt]{report}
\usepackage{iiscthes}
\usepackage{amsmath}
\usepackage{amsfonts}
\usepackage{charter}
\usepackage{amsthm}
\usepackage{amssymb}
\usepackage{graphicx,multirow}
\usepackage{subcaption}
\usepackage{subfiles}
\usepackage{url}
\usepackage{soul}
\usepackage{afterpage}

\newtheorem{theorem}{Theorem}[section]
\newtheorem{definition}{Definition}[section]
\begin{document}

\begin{frontmatter}
%
%

\title{Design of Trusted Market Platforms using Permissioned Blockchains and Game Theory}
	
\author{Shivika Narang}
\submitdate{MAY 2018}
\dept{Computer Science and Automation}
\degreein{Computer Science and Engineering}
\mscengg
\iisclogotrue 
\tablespagetrue 
\maketitle
\blankpage
\begin{dedication}
\begin{center}
DEDICATED TO \\[2em]
{\large \em My Parents and My Grandmother}\\
for\\
{\large\em Their Unending Love and Support }
\end{center}
\end{dedication}
\blankpage
\acknowledgements

My first and foremost gratitude is towards my advisor, Prof Y Narahari. He took me as a student despite my lack of understanding of game theory and proved to be the most understanding of advisors. In the last two years, he has been encouraging, kind, and supportive along each and every turn. His patience and kindness are virtues I hope to inculcate within myself. It was also thanks to him that I got an opportunity to spend a summer at IBM research where the work presented in this thesis began.

I am hugely indebted to Pankaj and Vinayaka who introduced me to the field of blockchains and bore with my shortcomings with grace. None of the work presented in this thesis would have been possible without their guidance. Many thanks to Pankaj for putting up with my incessant questions all through the summer, despite his busy schedule. A big thanks to Megha, for her technical contributions as well as her emotional support in the last year. This work would have been incomplete without her presence, patience, and warmth. Despite her many other obligations, she enthusiastically helped me at every step. I'm very grateful to have met someone with whom my thinking and tastes match. I hope we can continue our many adventures as long as possible.

My biggest supporters are my parents. Without them, their love, and their support, none of the things I consider my achievements would have been possible. Thank you so much for putting up with my rants and tirades all these years, for always giving me the freedom to dream, for helping me get centered on the many occasions I lost my cool. Thank you for teaching me to be self-reliant and independent. You both were my first teachers, and, all my accomplishments are a result of your teachings.

A lot of people helped make this thesis, and I would like to take this opportunity to thank all of them. Shweta and Ganesh, you both have been helping me out since the first version of this work was written up. Thank you so much for taking out the time from your busy schedules to guide me and my clumsy writing. A big thank you to Mayank and Vishakha for working even after their exams were over and helping proofread this thesis. Thanks to Arpita for her early inputs on this work and all the encouragement. A big thank you to everybody who had to sit through my mock presentations. I know they were nothing short of torture, but thank you for being so gracious.

Emotional support goes a long way in bringing a work like this to fruition. So a big thank you to Megha, Pooja, and Urvashi for being there for me, listening to my many rants, cheering me up, and going along with my crazy plans. Pooja and Urvashi, our tastes may not match in some things but thank you for going along with me anyways. Thank you guys also for being awesome company these last two years.

Lastly, but not the least, I'd like to thank my Nani. I wish you were here with me today. You are as much a parent to me as Mama and Papa. You raised me for thirteen years, and gave me the courage to believe in myself. I miss you every single day, especially on such occasions, which I want to share with you. Thank you for your love and warmth. I hope wherever you are, you're happy.



\begin{abstract}
The {\em blockchain\/} concept  forms the  backbone of a new wave technology that promises to be deployed 
extensively in a wide variety of industrial and societal applications. Governments, financial institutions, banks, industrial supply chains, service companies, and even educational institutions and hospitals are investing in a substantial manner in the hope of improving business efficiency and operational robustness through deployment of blockchain technology. This thesis work is concerned with designing trustworthy business-to-business (B2B) market platforms drawing upon blockchain technology and game theory. 

The proposed platform is built upon three key ideas. First, we use permissioned blockchains with smart contracts as a technically sound approach for building the B2B platform. The blockchain deploys  smart contracts that govern the interactions of enterprise buyers and sellers. Second, the smart contracts are designed using a rigorous analysis of a repeated game model of the strategic interactions between buyers and sellers. We show that such smart contracts induce honest behavior from  buyers and sellers. Third, we embed cryptographic regulation protocols into the permissioned blockchain to ensure that business sensitive information is not revealed to the competitors. We believe our work is an important step in the direction of building a powerful B2B platform that maximizes social welfare and enables trusted collaboration between strategic enterprise agents.
	
\end{abstract}
\blankpage
\makecontents

\notations

\begin{table}[h!]
\large
\hspace{0.1\linewidth}
\begin{tabular}{cc}
\hline
Acronym & Expansion\\
\hline
B2B & Business-to-Business \\
B2C & Business-to-Customer \\
CPA & Chosen Plaintext Attack\\
EDI & Electronic Data Exchange\\
MPC & Multi-Party Computation\\
MSNE & Mixed Strategy Nash Equilibrium \\
PSNE & Pure Strategy Nash Equilibrium \\
SE & Software Entity \\
SHA & Secure Hash Algorithm \\
SGPE & Subgame Perfect Equilibrium \\
\hline
\end{tabular}
\caption{List of Acronyms}
\end{table}
\blankpage
\end{frontmatter}
\chapter{Introduction}
\begin{quote}
{\em This chapter provides motivation and background for the work presented in this thesis, along with a brief review of relevant work. Platforms that enable collaboration between enterprise agents are few with limited functionality as a result of technological limitations. Relevant literature in economics as well as supply chains suggests that collaboration is needed in order to sustain high-quality offerings from sellers.}
\end{quote}
\section{Motivation and Background } 

A Business-to-Business (B2B) platform, as depicted in Figure 1.1, is one which enables interactions amongst enterprise buyers and sellers along with service providers such as logistics providers. Such a platform also serves the purpose of enabling information exchange amongst competing manufacturers/suppliers. B2B platforms are growing fast and intend to help upcoming businesses manage and streamline their work \cite{BtoB}. The design of robust B2B collaboration platforms and incentivizing cooperation on them constitute important research directions.

\begin{figure}[h!]
\label{fig:btob}
\hspace{0.2\linewidth}
\includegraphics[width=0.6\linewidth]{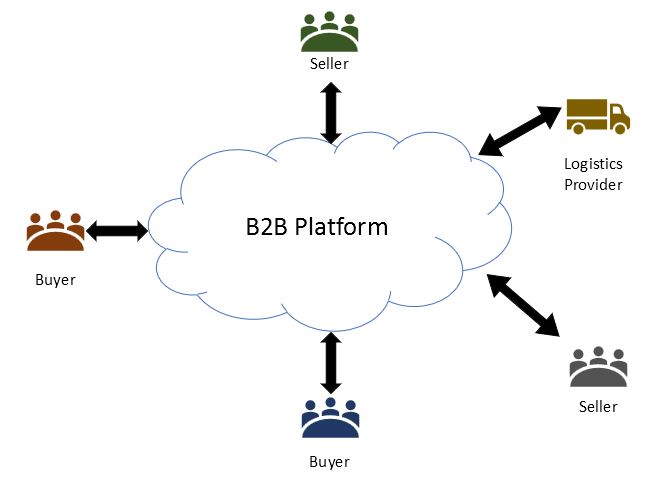}
\caption{A B2B Platform}
\end{figure}

A B2B procurement market platform is a B2B platform which helps facilitate the procurement process of enterprises. Businesses can utilize such platforms to trade goods and services which will be used to manufacture a good or deliver a service. 
While recent years have seen a lot of progress in Business-to-Customers (B2C) market platforms, especially with the advent of mobile applications, B2B platforms have not progressed technologically. They continue to be centralized and use point-to-point messages. One reason for this is a lack of technology which preserves privacy of the agents interacting on the platform, especially from the platform itself. While it need not be concerning for an individual if Amazon knows her purchase history, for an enterprise such information is business-critical. Consequently, the functionality that can be obtained by an enterprise from current B2B platforms gets limited. The work presented in this thesis aims to help solve this problem by proposing a blockchain-based solution. 

Recent progress in distributed ledgering technology, which is collectively referred to as {\em blockchain technology} has gained widespread attention from media and industry. In the last few years, it has quickly transitioned from an exciting technology, with a cult user base for the Bitcoin cryptocurrency \cite{NAKAMOTO2008}, to being seen as a powerful tool for transformation of business ecosystems. The technology allows users on a distributed peer-to-peer network to maintain a distributed ledger without any need for a centralized intermediary. At the core of any blockchain platform are,
\begin{enumerate}
\item A mechanism for secure and distributed consensus protocol for maintaining the ledger, and
\item A framework to autonomously execute {\em smart contract}s based on the state of the distributed ledger.
\end{enumerate}

\noindent The initial research focus on blockchain has been on various security, privacy, and game theoretic aspects of distributed consensus mechanisms for applications in cryptocurrencies. The focus of our work is on identifying and providing solutions to the issues that are critical to making blockchain a tool for transforming business ecosystems. 

An advantage of a blockchain-based market platform is that rules can be imposed on the transactions that occur on the platform, by way of smart contracts. For instance, smart contracts can be used to govern the prices charged, and can be a means of ensuring that goods are traded at a fair price. The rules that are to be enforced can be decided jointly after negotiations between the buyers association and the sellers association. The appropriate form of blockchain technology for such a purpose would be permissioned blockchains, where the access to the data is restricted to users of a company or consortium of companies that manage the blockchain. Permissioned networks are essential for business-to-business (B2B) collaborations as businesses need to know who they are transacting with and there is a need to control the purposes for which they transact with different entities. 

\noindent a typical B2C market platform, such as Amazon or eBay, when a buyer decides to buy a particular good, she typically searches for it and is able to view different vendors who sell the good. She then proceeds to buy from the vendor who she believes offers the best tradeoff between quality and price. 
Studies indicate that to induce sellers to consistently provide high-quality products, it is necessary to make it publicly known when a seller fails to do so. B2C market platforms achieve this by way of ratings for each good sold by a vendor. 

Our work aims to enable an enterprise buyer to do the same on a B2B market platform, while being incentivized to be honest. 
Further, we use a repeated game formulation to study various pricing rules and punishment strategies, so as to identify the ones that best incentivize high-quality offerings from sellers. This analysis is intended to help design appropriate smart contracts which will define the interactions that occur on the platform. 


\section{Review of Relevant Work} 

A motivation of our work is to achieve sustained cooperation amongst buyers and sellers. Buyer-Seller interactions over market networks have been studied from various aspects.  From an economics point of view Fainmesser found that the frequency of high-quality given by a seller in large market networks increases with moderate and balanced competition, and that the inclusion of institutions like litigation can sustain cooperation in denser networks \cite{fainmesser2012community}.  

Zhang and van der Scharr study this problem from a crowdsourcing aspect. They consider a mechanism, where each worker has reputation in the set $\{ 0,1,..., K_K \}$. After each period, based on the report by the requester, assumed to be truthful, the reputation is either incremented or decremented. If the reputation of a worker falls below a threshold, she is put into isolation. They study how the choice of the threshold affects cooperation. Conditions are given on $K_K$ as a function of the system parameters on when the protocol will have the workers putting in high effort to take on the tasks given to them, as a Nash equilibrium.

In order to sustain high-quality from sellers in a market, information about how an agent behaves with other agents is needed\cite{wolitzky2015communication}. However, this information often remains local, known only to the two parties and perhaps their neighbors. Wolitzky \cite{wolitzky2015communication} studies two communication technologies to replicate public information. 

While the problem at hand is of market interactions, the broader motivation of our work is aimed towards supply chains. Much like in market networks, information sharing in supply chains has been found to reduce costs both theoretically and experimentally \cite{cachon2000supply,ganesh2014distribution,gazzale2011remain}, much like in market networks.

We enable information sharing, and hence collaboration, between enterprise agents by providing a cryptographic protocol for computing a ratings vector for the sellers, without revealing their own input. This problem of jointly computing reputation of agents or building reputation systems is neither new nor trivial. Procaccia et al  \cite{procaccia2007gossip} use a gossip-based aggregation technique to provide a simple decentralized reputation system without a need for complex data structures. Jiang \cite{jiang2013towards} attempts to design reputation systems that are robust to attacks such as malicious ratings. Haghpanah  \cite{haghpanah2011trust} focuses on reputation systems for supply chains which do not downweight the value of feedback given by other participants. The method proposed in this thesis, for rating sellers on a B2B platform, is able to achieve the salient features of all of the three papers mentioned above. Our problem also fits into the broader area of collaborative rating, which is an important problem for recommender systems \cite{ducollaborative}. In the problem of collaborative rating, ratings must be computed for a set of $m$ objects jointly by $n$ users. Each user provides ratings on a strict subset of the $m$ objects, and complete ratings must be computed from these partial ratings. The problem of computing collaborative ratings itself is a specific case of the matrix completion problem, pioneered by Candes and Retch \cite{candes2009exact}.

\section{Contributions of the Thesis} 
Having established the positioning of our work, we now provide an outline of the specific contributions of the work presented in this thesis. Figure 1.2 depicts the various building blocks of the proposed platform and motivates the various contributions of this work. \\[2mm]

\begin{figure}[h!]
\begin{center}
\label{fig:cont}
\hspace{0.1\linewidth}
\includegraphics[width=0.99\linewidth]{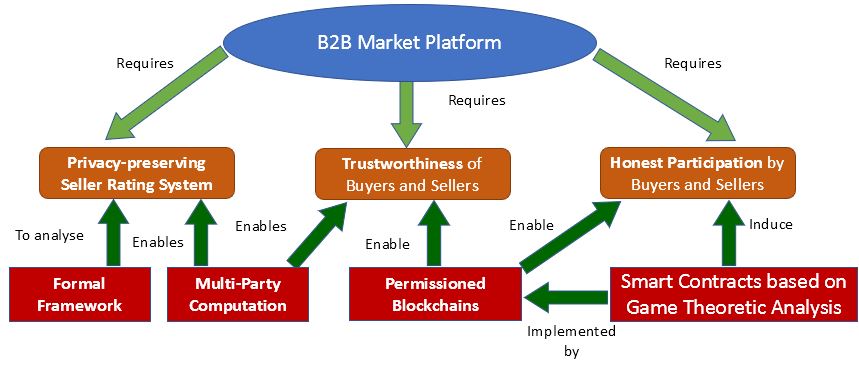}
\caption{Building blocks of the proposed blockchain-based B2B platform}
\end{center}
\end{figure}

\noindent
\subsection*{A Formal Framework for Secure B2B Collaboration} One of the key aspects of B2B collaboration is that, often, the cooperating entities are also competing entities and the information required for cooperation is business sensitive information from competitive perspective. As discussed, designing a platform that successfully enables privacy-preserving B2B collaboration is a non-trivial task and requires a formal discussion. Therefore, there is a need for a framework to formulate the requirements of a secure and private B2B collaboration. We provide a formal notation for studying and analyzing all the components of a B2B platform and the events that occur on the platform. To the best of our knowledge, we present the first such framework for B2B collaboration.

\subsection*{Game Theoretic Modeling and Analysis} We use repeated games to analyze the strategic interactions between buyers and sellers in the context of a supplier rating problem. We compute the equilibrium behavior under different strategies and pricing settings. The analysis reveals that the equilibrium attained is contingent on the sellers' discounting factor for future payoffs, as well as, on the profit margin of the pricing system. We study markets where social welfare is maximized when sellers provide {\em high\/} quality and deduce conditions under which high quality offerings constitute an equilibrium. These characterizations are of independent interest in themselves. There is prior literature on game theoretic analysis of mining in permissionless networks, exemplified by \cite{kiayias2016blockchain,lewenberg2015bitcoin}. To the best of our knowledge, there is no prior work on a framework to study game theoretic aspects of blockchain based B2B collaboration. 

\subsection*{Implementation using Permissioned Blockchains and Cryptographic Protocols} We show how to employ permissioned blockchain networks to securely and privately mimic an oracle's solution to the supplier rating problem. We deploy a smart contract that implements the business logic in a given mechanism with the help of two regulation protocols: (a) a public perception protocol and (b) a monitoring protocol. The public perception protocol is a cryptographic protocol that aggregates buyer feedback about sellers into a rating, known as the public perception vector, while preserving privacy. The purpose of the monitoring protocol is to disincentivize dishonest feedback by the buyers. We show that our implementation is a secure and private simulation of the oracle's supplier rating.


\newpage
\section{Outline of the Thesis}

The organization of this thesis is as follows.
\subsection*{Chapter 2: Foundations of Blockchain Technology} This chapter provides the foundations of blockchain technology. Section 2.1 covers the building blocks of blockchain technology. It begins by explaining the different contexts in which the term {\em blockchain} is used, followed by details of the blockchain data structure. We describe the organization of data in the blockchain, followed by how blocks are added to a blockchain, introducing two commonly used mining techniques. Section 2.2 discusses the key features of blockchain technology, its role as an asset ledger and introduces smart contracts. The chapter concludes with Section 2.3 which gives a discussion of a commonly used classification of blockchains into permissionless and permissioned blockchains. We discuss the issues that arise with each of them like mining and scalability. We describe the various types of permissioning that may be employed.

\subsection*{Chapter 3: Problem Formulation and Approach} The third chapter of the thesis presents the formulation of the problem studied in this thesis along with the approaches followed. The chapter begins with Section 3.1 which first gives a brief description of the technology being used to deploy current B2B platforms. It discusses the limitations of the current approach, and motivates the use of permissioned blockchains for designing a B2B platform. The section then concludes with the requirements of such a B2B platform. Section 3.2 then defines a formal framework to study interactions on the platform. To the best of our knowledge, this is the first attempt at formally defining interactions over a blockchain network. 

Section 3.3 goes on to detail the building blocks of this framework. It begins with a description of the model, defining the repeated interactions as a repeated game. Following this discussion, we  introduce the  cryptographic protocols that are used to compute the ratings and the rating system adopted. The chapter concludes with the definition of the public perception vector, that is we define how the feedback from the buyers is aggregated into a ratings vector for the sellers.

\subsection*{Chapter 4: Game Theory Based Smart Contracts} The fourth chapter is devoted to game theoretic analysis of the problem presented in the third chapter. The chapter begins with a few preliminaries on game theory. Section 4.1 formally defines strategic form games and Nash equilibrium. This chapter also discusses some examples of strategic form games and the Nash equilibria that arise in each of these games. Section 4.2 goes on to define Extensive Form Games, and what a Nash equilibrium for an extensive form game is. It also presents a stronger notion of equilibrium called the subgame perfect equilibrium. Section 4.3 concludes the game theory preliminaries by defining repeated games and how subgame perfect equilibria can be characterized for a repeated game. It also explains the punishment aspect of the problem studied in the thesis. 

Section 4.4 begins the game theoretic analysis of the problem. This section considers a pricing rule called ``Homogeneous Pricing'', where all sellers sell at the same price. It is shown that the only pure strategy equilibria for such a system can be either giving only high-quality or only low-quality. This section studies two different punishment models, one where sellers are punished by individual buyers for providing low-quality in a particular round. In the other model, a seller is punished collectively by the all the buyers in the market if his rating drops below a threshold value. It is also shown that the latter punishment model is better at incentivizing high-quality offerings from sellers. 

Section 4.5 considers a setting where sellers can sell their products at one of two prices. Within this setting two cases are considered, one where the price at which a seller sells never changes, and one where the price at which the seller sells is determined by his public perception. It is found that both these settings reduce to different cases of homogeneous pricing, contingent on the parameters of the system. Section 4.6 takes the idea of the price being determined by the seller's rating forward and the seller's rating is linearly transformed to determine the price which he charges. This section finds behaviour that was not observed in the previous two sections. The chapter concludes with the observation that the rather complicated behaviour observed in the case of continuous pricing, prevents a closed form expression for characterizing the Nash equilibria in this setting.

\subsection*{Chapter 5: Cryptographic Regulation Protocols} This chapter presents details of the cryptographic protocols deployed on the platform which ensure that the requirements of the B2B platform listed in Chapter 2 are met. We first present some relevant preliminaries on public-key cryptography and secure multi-party computation required for the protocol, such as homomorphic encryption. This is covered in Section 5.1. Sections 5.2 and 5.3 detail the two subprotocols, namely the Monitoring Protocol and the Public Perception Protocol. The former ensures that buyers are honest in their feedback, and the latter is responsible for actually aggregating buyer feedback into a ratings vector for the sellers. The chapter concludes with a discussion on the properties of these protocols and how the appropriate requirements are met by means of these protocols.

\subsection*{Chapter 6: Summary and Future Work} This chapter presents a summary of the thesis, along with some directions possible for future work. Section 6.1 presents a summary of the results presented in the preceding chapters. Section 6.2 discusses possible variations in the model presented in this thesis, and accordingly lists the scope of future work in this problem. 

\section{Conclusion}
A review of existing literature shows that to ensure consistent quality is received from a seller, information of their behaviour should be made public to all the buyers. While most B2C platforms allow for buyers to rate sellers,  progress has been stalled in the development of platforms enabling B2B collaboration. One of the reasons for this is that interactions on such platforms should be such that privacy of the agents is preserved. 

A B2B market platform requires a privacy-preserving way to rate the enterprise sellers or suppliers. The rating system must be robust to attacks such as malicious ratings attacks. This problem is solved using permissioned blockchains and Multi-Party Computation protocols. The other requirement of a B2B market platform is to incentivize sellers to provide high-quality. We conduct game theoretic modeling and analysis towards this end, which will be used to design smart contracts on the blockchain.

In the following chapter, we provide an overview of blockchain technology that is relevant to our work.
\chapter{Foundations of Blockchain Technology}
\begin{quote}
{\em This chapter introduces the concept of blockchains along with several issues and applications associated with blockchains. The term blockchain can refer to the blockchain network, the distributed ledger or the data structure, which makes the contents immutable. As a result of immutability, the use of blockchains in a trustless setting makes the problem of adding new blocks non-trivial. Thus, mining of new blocks becomes a natural concern when the blockchain network is one without any restrictions. However the issues associated with the deployment of blockchains change when there is a restriction on the roles a user can take up. This chapter discusses the aforementioned aspects in detail and concludes with a glance at the application space of blockchains. 
The contents of this chapter are culled out from a recently published survey paper on blockchains \cite{narang2018}.}
\end{quote}

\noindent {\em Blockchain technology} is promising to herald a new revolution in a rich variety of industrial and societal settings. Many important global businesses are trying to embrace the blockchain technology to find solutions to their difficult problems. Governments, financial institutions, banks, industrial supply chains, service companies, and even educational institutions and hospitals are investing substantial sums of money in the hope of improving business efficiency and operational robustness. 

Under the pseudonym {\em Satoshi Nakamoto\/}, an anonymous  person or a group of persons wrote a brilliant paper \cite{NAKAMOTO2008} in 2008 in which the bitcoin digital currency was introduced. This paper introduced a simple but powerful data structure which Satoshi Nakamoto  called the {\em blockchain\/}. Using this data structure, which uses hashing to achieve immutability of data, and through an intelligent design of incentives. This paper launched the bitcoin revolution, which has now grown to the point that bitcoin is now a household topic. Blockchains are now used in many different applications apart from cryptocurrencies. With the functionality of a scripting language accompanying the blockchain to build {\em smart contracts}, introduced by Ethereum\cite{BUTERIN2013}, decentralized applications or Dapps were launched and have escalated far more quickly than anticipated. The term {\em blockchain} now goes beyond particular applications like Bitcoin and Ethereum, and is used to signify the data structure, the data repository, which records the transactions, as well as the network on which these transactions occur. We now explore each of these.

\section{Blockchain Building Blocks}
A blockchain network is typically a peer-to-peer network where each node has computing resources and a data repository. The data repository at each node contains an  an up-to-date copy of the blockchain. Figure 2.1 illustrates a blockchain network and its essential components: a computing resource and data repository within each node and a lack of centralization.  

\begin{figure}[h!]
\label{fig:one}
\hspace{0.1\linewidth}
\includegraphics[width=0.8\linewidth]{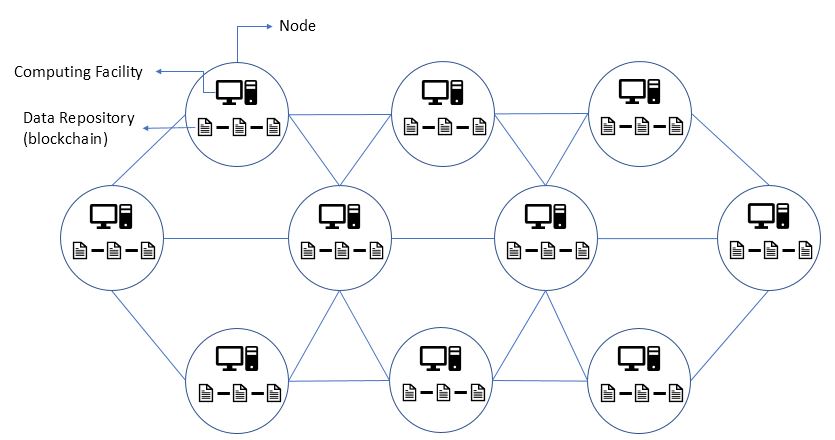}
\caption{A Blockchain network}
\end{figure}

\subsection{Blockchain Data Structure}
The blockchain data structure is a structured collection or shared ledger of blocks. Figure 2 depicts a part of a typical blockchain. In this subsection, we describe the data structure that is used in the bitcoin blockchain, to bring out all essential features of a blockchain. A blockchain may contain any number of blocks. For example, the bitcoin blockchain (as of January 2018) contains more than 210000 blocks. 

\begin{figure}[h!]
\label{fig:three}
\hspace{0.1\linewidth}
\includegraphics[width=0.8\linewidth]{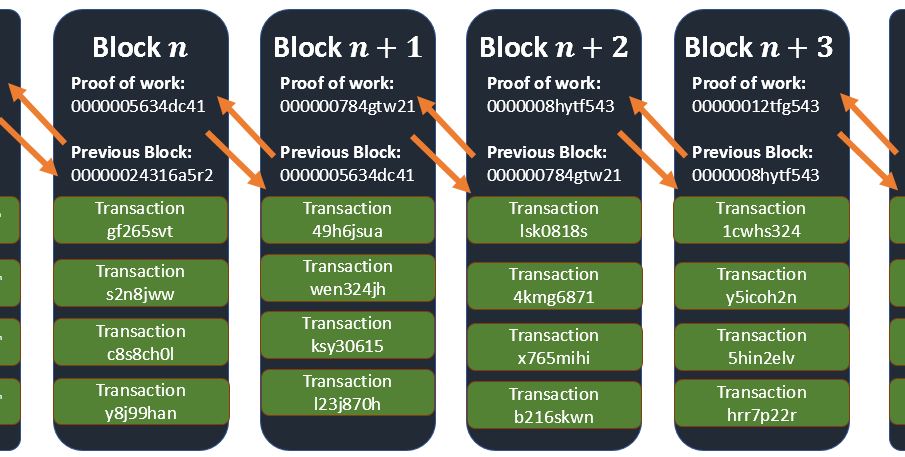}
\caption{A Blockchain}
\end{figure}

As a data structure, a blockchain is a doubly linked list of the blocks. Thus, each block has a pointer to the next block as well as a pointer to the previous block.  Each block contains the following:
\begin{itemize}
\item Data corresponding to a number of transactions. Each transaction may involve a small subset of nodes and is represented in some digital form. A block may contain any number of transactions. For example, a block in the bitcoin blockchain could contain hundreds of transactions.
\item A hash value corresponding to the current block.
\item A hash value corresponding to the previous block. 
\end{itemize}
Each hash value above is also called {\em proof-of-work} for that block. Proof-of-work for a block is computed by solving a difficult cryptographic puzzle: this puzzle takes as input the transaction data of that block and produces a final hash value, after a very large number of hash computations, such that the final value that satisfies a certain  criterion that is difficult to achieve (for example, the hash value should start with a string of four zeros, etc.). Because of the intensive nature of this computation and the massive amount of computational work involved in producing this final hash value, the latter is aptly called {\em proof-of-work\/}.

Since every block contains proof of work for the current block and proof-of-work for the previous block, the blockchain becomes virtually immutable. This is because, in order to modify a block, not only do you have to modify the current block but also all successive blocks.

\begin{figure}[h!]
\label{fig:two}
\hspace{0.05\linewidth}
\includegraphics[width=0.9\linewidth]{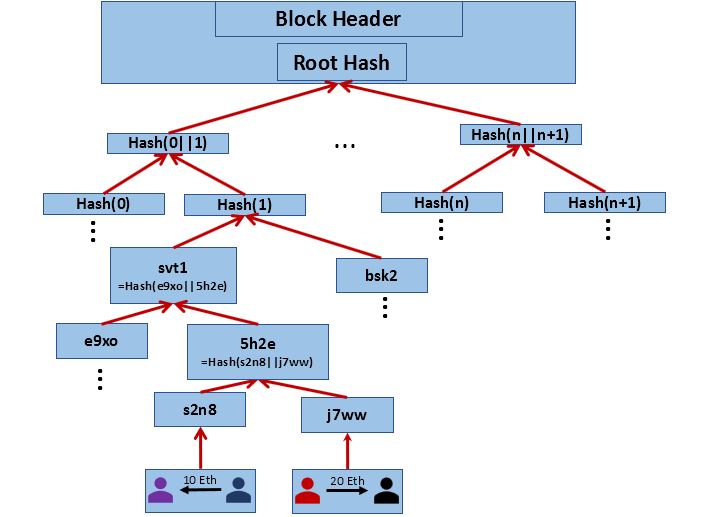}
\caption{Merkle Tree}
\end{figure}

The data corresponding to each block is organized as a Merkle tree as shown in Figure 3. Merkle tree is a binary tree data structure where hashing is used to facilitate efficient verification of the contents of the tree. The leaf nodes contain the hash values of individual transaction data. Each internal node contains only a hash value that is computed using the hash values contained in its children. The root node will thus contain a hash value which is called the root hash value. Given a Merkle tree, in order to verify that a given a set of transactions is precisely the same as the transactions stored in the Merkle tree, we only have to
compute the hash value from the given transaction data and verify that the hash value is the same that of the root hash value in the given Merkle tree. A new block is included into the blockchain using a process called {\em mining\/} which is described in the next section. Once a new block gets included, the updated blockchain becomes available at all the nodes. The manner in which the blockchain data structure is organized, updated, and replicated ensures that the blockchain is  a shared, synchronized, immutable, distributed ledger  created through distributed consensus; all nodes have access to the ledger and can verify the truth. Simply speaking, the blockchain elegantly implements {\em tamper-proof record keeping\/}. The blockchain protocol is decentralized and eliminates the need for intermediation by trusted third parties. Blockchain technology thus represents a deep one that combines cryptography, data structuring, and distributed consensus in a brilliant way.

Blockchain technology provides a way to implement a distributed, shared ledger without the need for a central trusted entity or trusted third parties. There are other ways to implement a distributed, shared ledger but blockchain technology is superior to the existing approaches because of the above features. For any immutable data structure, which continues to be decentralized, the issue of adding new blocks to the chain is not  trivial. We now briefly study common techniques of adding new blocks to the blockchain.

\subsection{Mining of Blocks}
The addition of a new block to  a blockchain is based on inverting a hash function, whose computational complexity is believed to grow exponentially in the length of the input. This process is  known as {\em mining\/} or {\em validation\/}. Mining secures the blockchain system from fraudulent transactions. Mining can be done by any node in the (public) blockchain. The task of a miner is two fold: (1) to validate the data in the block and (2) to append the valid block to the already existing blockchain. Note that the success of the blockchain is attributed to the way mining is done i.e. how the verification of data is done in a decentralized, anonymous fashion. In the next two subsections, we present  how this is achieved via the standard process called  proof-of-work and a modified process called proof-of-stake. 	

This section assumes a bitcoin/ethereum type of (public) blockchain. However, the principles are relevant for other types of blockchains as well. 
\subsubsection{Proof-of-Work}
The key idea behind proof-of-work is to make the selection of a miner in a way that nobody can monopolize the system by creating fake nodes.  This is ensured in proof-of-work by making the nodes to compete to mine the block using their computation power. This would imply that no single node  can monopolize the blockchain protocol as buying that much computation power would be next to impossible. 

The proof-of-work is based on inverting a hash function where the hardness of the computation can be controlled. This cryptographic computation is a hash function based computation where each miner is required to find a random numeric value {\em nonce} such that the cryptographic hash function of the nonce when combined with the data and the hash value of the previous block results in a hash  value that is less then a specific target value. Specifically, $$H(\textit{nonce}||\textit{hash-of-previous-block}||\textit{data-of-current-block}) <\textit{ target-value}.$$
The $\textit{target-value}$ is adjusted automatically between the nodes in the blockchain so as to maintain the difficulty of mining to be a certain time duration. Note that once a miner broadcasts a mined block, every other node can verify that the mined block is valid and thus will append this block to its own blockchain. The proof-of-work technique described above prevents the problem of double-spending the same currency.

A disadvantage of proof-of-work is that it needs a massive amount of computation power to validate a single block. Another problem is that mining has become rather proprietary hardware centric  which leads to centralization issues as people who are manufacturing such hardware can control the blockchain system. Therefore, there is a shift to other mining techniques like proof-of-stake.

\subsubsection{Proof-of-Stake}
In proof-of-stake, nodes are selected for mining in proportion to the blockchain currency each node is holding as opposed to the computation power in proof-of-work. In proof-of-stake protocol, any node which holds blockchain based currency can become a miner by depositing its currency to the blockchain. Selecting the validator can be done in several ways, such as {\em Randomized block selection} and {\em Coin age-based selection}.

The proof-of-stake protocol is significantly more computationally efficient over proof-of-work protocol in terms of energy consumption, however, this protocol suffers from other disadvantages like nothing at stake and long range attacks.

Immutability and mining protocols to achieve consensus contribute towards a blockchain being a trusted ledger. We now explore the key features of blockchain technology which are a consequence of the building blocks discussed in this section.
\section{Blockchain Technology: Key Features}

The unique value proposition of blockchains is that they enable the creation of digital solutions with built in economics. This capability traces its roots to the first applications of blockchains: bitcoins as a digital currency. A necessary condition for a digital currency is the existence of a trusted ledger which can maintain the balances of the account holders. Nobody apart from the account owner should be able to reduce the account balance; not even the entity responsible for maintaining the ledger. This is precisely what is achieved by the blockchain. As a result, today blockchains are being used to develop applications which require decentralized asset registries, such as land registry, tamper-proof academic transcripts and even a registry for diamonds which tracks a diamond right from when it is mined\cite{CROSBY2015}.

\subsection{Asset Ledger}
At its very core, a blockchain is a trusted ledger. As explained in the previous section, the technological innovation of blockchains is that the trust in the ledger is ensured using a combination of cryptography and distributed consensus. The most important consequence of using a shared, trusted ledger is that everyone involved is able to agree on a single version of the truth. Thus, an immediate application of blockchains becomes as an asset registry. Assets such as land and property are the most common assets used as collateral to secure a loan, and thus require a trusted, immutable registry.  


One interesting question in creating a blockchain based asset registry relates to privacy. Does storing asset ownership on a public registry imply that everyone knows who owns what? Public blockchains like Ethereum and Bitcoin blockchains solve this system by using pseudo-anonymity. Every agent in the system is recognized by its account address (public key) but the blockchain does not store a mapping between this account address and the real-world identity of the agent. With these blockchains, thus, the mapping of assets to account addresses is common knowledge but the mapping between the account address and real-world identity may be kept private (known only to the individual agent). In the context of blockchains being used as an asset registry for cryptocurrencies, it is this pseudo-anonymity property which poses challenges to governments and law enforcement agencies worldwide. 

With permissioned blockchains (discussed in detail in the next section), the issue is less pronounced but still exists. With permissioned blockchains who all can become a part of the blockchain network is controlled by an administrator, so information sharing (of the asset registry) is within a group. Often, however, there may be cases where two (or more) agents in the network may want to undertake a transaction privately. Here again is the requirement for privacy. Interesting solutions to this challenge continue to be explored:  cryptographic techniques like zero-knowledge proofs, ring signatures, and hash functions can be combined in interesting ways to achieve privacy while still ensuring trust in the underlying ledger.

\subsection{Smart Contracts}
Powerful as they are, the ideas of a trusted asset registry and a cryptocurrency are only part of the complete picture. It is when these two ideas are combined with the idea of self-enforcing contracts known as {\em smart contracts\/} that the real power of blockchains becomes apparent. Smart contracts are the building blocks to create any decentralized application.

A smart contract can be thought of as a conditional transaction, that is, a transaction (or a group of transactions) which executes when certain conditions are met. The conditions required for the transaction are written in a high level programming language like Solidity, the scripting language accompanying Ethereum \cite{BUTERIN2013}. In this interpretation, a smart contract is a conditional transaction of value from one agent (account) to other agent(s) conditional on certain events that may occur in the future. Once the contract is \textit{deployed} on the blockchain, it is guaranteed to execute as per the conditions specified. This ``guarantee'' stems from the distributed nature of the blockchain. When a smart contract is deployed on the blockchain, a copy of this smart contract resides on every single node in the underlying peer-to-peer network. Each node executes the smart contract code to check whether or not the conditions specified in the contract have been met or not. The verification of these conditions is subject to the same consensus algorithms that a transaction is, thus ensuring that the smart contract executes if and only if the nodes in the blockchain agree that the conditions have been met. To summarize, smart contracts are self-executing contracts written in a high level programming language and deployed on a decentralized peer-to-peer network.

Traditionally, contracts underlie a vast span of global commerce, for example,  procurement contracts detail the requirements and the conditions under which payments would be made or penalties imposed. Commercial contracts build up the trust for the two (or more) parties thus facilitating trade across the globe. In the blockchain world, these contracts are encoded in a programming language. In this context, smart contracts are computer software which implement commercial contract using cryptography and distributed systems. The great advantage of doing so is that once deployed, these smart contracts are guaranteed to execute and do so automatically. Moreover, since written on a blockchain, when transactions happen with the help of smart contracts, these transactions are immutable. Not only does this enhances trust in the contract, it also reduces manual overhead of contract verification, payments processing etc.  

To summarize, the key properties of smart contracts are:
\begin{itemize}
\item \textbf{Decentralized:} The smart contracts are executed on hundreds of computers without being controlled by any human. Thus, there are fewer risks of manipulation or fraud.
\item \textbf{Autonomous:} The smart contracts are executed by themselves once they are launched and thus they speed up the trading or the process for what the smart contracts are designed for.
\item \textbf{Auto-sufficiency:} The smart contracts can be sufficient in their own way e.g. they can collect money, provide automated access to storage units, vehicles, make payment for insurance or rent, distribute resources, etc.
\end{itemize}




\noindent Trusted ledgers and conditional transactions are required not just by platforms like Ethereum, where anybody can join, but also in more private settings. We now explore blockchains in both public and private settings, and the associated issues.

\section{Blockchain Types: Permissionless and Permissioned}
A common way to categorizing the types of blockchains is based on the permissioning used. Permissioning refers to the requirement of permission to be able to take a view or add blocks on the blockchain. Permissioning is a natural requirement for several applications where there is a requirement of confidentiality. In this section, we cover permissionless and permissioned blockchains, along with the consequences of the permissioning model chosen. 

\subsection{Permissionless Blockchains}
Permissionless or public blockchains are essentially blockchains that do not require permission to view, join, or mine blocks. The lack of permission, may be essential to the application, as in the case of some of the most popular blockchain applications such as Ethereum and Namecoin, a decentralized platform to issue domain names. Typically, such blockchains are intended for use by the masses. Several issues come up, or are foreseen, as the majority of blockchain projects are yet to reach a level of maturity, in the deployment of public blockchains. We now discuss some of these issues.

\subsubsection{Mining} 
Allowing users to freely join the network also necessitates much stricter security. This is required as public platforms cannot assume that nodes will not be malicious. The only underlying assumption critical to guarantee the correct execution of such platforms, is that the majority is honest. As a consequence, most common permissionless blockchain platforms use proof-of-work mining, which typically takes longer to validate a block. Ethereum, which has a relatively short validation time, when compared to other public blockchains, typically takes 12 seconds to mine a block. On the other hand, proof-of-stake mining, such as that which is used in Hyperledger, ensures that transactions are validated in 0.5 seconds. 

\subsubsection{Scalability} 
Zohar studies the problem of scalability as it pertains to cryptocurrencies, which are typically implemented on public blockchains \cite{zohar2017securing}. Currently, there are set limits for the size of a block that can be put on the blockchain, which limits the number of transactions that may be put on the block. With these limits held constant, as the number of users grows on the blockchain, the number of transactions also increases, thus leading to further delays in the time required to mine a block. Further, the need for computational power for such platforms is expected to grow much faster than the growth in the available computational power, thus leading to centralization in the validation of blocks \cite{PublicBC}. Needless to say, this goes against the very purpose of a decentralized ledger. Thus, permissionless blockchains face a major challenge of scalability. 
Sompolinsky et al \cite{sompolinsky2016spectre} offer solutions to the scalability problem by proposing an alternate consensus scheme, built on Nakamoto's work \cite{NAKAMOTO2008}.

\subsubsection{Privacy} While privacy is often a concern cited to justify the need for a private network, for an individual, a public network provides more privacy. As there is no restriction for joining the network, an individual may have multiple identities on the same network. Transactions on the network are typically attributed to public keys of the parties involved. Thus, by creating aliases on the network, individual users can each have multiple public keys and hence ensure their privacy. This privacy, however, is limited to being pseudonymity. There have been cases where leaks in transaction histories of merchant websites have led to revelation of the true identity of the user \cite{BreakAnon}. 

\subsubsection{Application Space} The use of permissionless blockchains is reduced to applications which are trustless. Allowing anybody to join the network obviates the assumption of trust, and with it, all applications which require a measure of trust. Consequently, the realm of cryptocurrencies has seen the most prominent deployment of public blockchains. However, applications of permissionless blockchains extend far beyond cryptocurrencies. Some applications include prediction markets, land registry, voting and even governmental services being provided through blockchains \cite{swan2015blockchain}. 

\subsection*{Ethereum}
Ethereum is a public blockchain platform, with an accompanying programming language, Solidity, to enable users to build distributed applications on it 
\cite{swan2015blockchain}.  
Ethereum provides rich  functionality to solve computational problems  through smart contracts. This provision can be used by developers to build apps on top of Ethereum. Currently deployed decentralized apps, or dapps, include those for voting, crowdfunding, online marketplaces, auctions, etc.

Ethereum was proposed in 2014 by Vitalik Buterin, a young programmer who, at that time, was known primarily for writing extensively on Bitcoin. It was an attempt at heralding in the use of smart contracts in the cryptocurrency domain, by providing a scripting language, to write smart contracts, in order to implement the intended business logic \cite{BUTERIN2013} \cite{WOOD2013}. 

\subsection{Permissioned Blockchains}

Permissioned blockchains are typically not open for users to join freely. However, the term is used rather broadly. Any blockchain where users need permission to take up a role are considered to be permissioned blockchains. These include applications like Monax, where every single possible action, such as adding new users, deploying smart contracts, requires explicit permission. On the other hand, applications such as Quorum, where the only permission that is required is for a node to be a validator, are also categorized as permissioned blockchains. 

An extreme case of a permissioned blockchain is a {\em private blockchain\/} which is centralized in that write permissions are given only to a highly trusted central entity;
read permissions are available to all participants. Only chosen players are permitted to join the network leading to a closed loop environment. Such blockchains are appropriate for handling certain employer-employee transactions in organizations.

A widely used classification of permissioning followed is as follows.
\begin{itemize}
\item Simple Permissioning: Typically, permission is only required to validate new blocks. With simple permissioning, new users may be registered by existing users, and smart contracts may be deployed by any of the users.
\item Complex Permissioning: Usually requires ``fine-grained'' permissions, including being able to view the content of the blockchain, adding new users, validating new blocks. Needless to say, the granularity of the permissions required, are a consequence of the business model and requirements of the use case.
\end{itemize}

\subsubsection{Privacy Models}
There is variation in the privacy provided by upcoming and currently deployed private blockchains. For any blockchain-based platform, the transactions that have been validated typically form the content of the blockchain, privacy of transactions is a part of the privacy being provided by the platform. Permissioned blockchains typically have three types of transactions:
\begin{itemize}
\item Public Transactions: Similar to transactions on public blockchains.They can be seen by everybody on the network, and thus, can be validated by any validator.
\item Private Transactions: Such transactions are visible only to the parties involved in the transaction. Consensus must be reached by all the parties involved for the transaction to be validated. A validator present on the network but not involved in the transaction cannot view, let alone validate the transaction.
\item Group Transactions: Transactions are visible to a subset of the nodes, chosen by the parties involved in the transaction. However, validators of the transaction are typically the parties involved in the transaction.
\end{itemize}

\noindent Different types of permissioned blockchain applications adopt some combinations of these three  types of transactions, based on their needs. Monax, while using complex permissioning, offers no privacy in transactions\cite{ComparisonBC}. All transactions are public transactions. Corda, implemented for secure communication in the financial sector, only has private transactions, however, parties involved in the transaction may allow regulator nodes to be involved in the consensus. Quorum is an application that provides for all three, public, private and group transactions.

\subsubsection{Consensus} While public blockchains require proof-of-work based consensus, which are computationally expensive, permissioned blockchains can afford to have much more computationally efficient consensus protocols, and further have varying provisions for privacy of transactions. As a result, there is some variance in the consensus algorithm used. While a large fraction of the permissioned blockchains currently deploy proof-of-stake mining, some, like Quorum, use consensus similar to proof-of-stake, but may not provide byzantine fault tolerance. Applications like Corda simply have private transactions, and the parties involved must agree on the transaction to reach consensus. 

\subsubsection{Application Space} Permissioned blockchains developed as a response to the needs of enterprises to have a private tamper-proof distributed ledger. Consequently, many of these applications are geared towards enterprise networks, such as the banking industry and supply-chain finance and transparency \cite{swan2015blockchain}. However, the applicability extends well beyond this realm, including decentralized identity, which aims to decentralize the traditionally centralized domain of storing potentially confidential records and letting users choose the who is allowed to view which records.  

\subsubsection{Hyperledger}
The {\em Hyperledger project} is a joint open-source effort led by a consortium of the Linux Foundation, towards the advancement of blockchain technology. The consortium currently has more than 50 member institutions and corporates that are interested in developing the capabilities of blockchain applications. The umbrella of the Hyperledger project, contains various projects within it, each with different flavors of blockchains.

\noindent{\em Hyperledger Fabric} is a permissioned blockchain framework, which can be implemented, according to the needs of the enterprise. Various components such as deploying the consensus protocol and membership services can be chosen to be included, or not, as per convenience \cite{cachin2016architecture}. The motivation behind Hyperledger Fabric is to enable enterprises and groups to deploy blockchains as per their own needs. 

\section{A Vast Canvas of Applications}
\label{sec:app}
We have seen that a well designed blockchain based platform, whether public, or private, or permissioned,  is guaranteed to be trustworthy, immutable, secure, and enables powerful smart contracts to self-execute in an automated way. In addition, a platform that is built upon a public blockchain is guaranteed to eliminate intermediation and enables sharing in a decentralized and democratic way. A platform based on a permissioned blockchain can also harness some of the key advantages of public blockchain based platforms. 

The numerous desirable characteristics of blockchain based systems have opened up a wide canvas of  powerful industrial and societal applications in the recent years. In the discussion below, we first briefly describe a number of such applications in different domains. Following that, we provide a more detailed treatment of four applications:
tamper-proof land registries, academic transcripts, crowdfunding, and a B2B supply chain platform.
This section is by no means exhaustive and the reader is urged to look into a large number of applications covered in the following
articles: 
\cite{CROSBY2015,DELOITEE2017,RBI2017,IANSITI2017,SPECTRUM2017,TAPSCOTT2016,UNDERWOOD2016}. 
There are many other sources in the literature and webpages which contain a  wealth of information. The references we have provided here cover only a small fraction of content available on the web and we apologize to the authors of a large number of articles we are unable to cite here.

Depending on the application, a certain type of blockchain will be more appropriate than others. We have already seen the differences among public, private, and permissioned blockchains; the nature of the application and the requirements mandated by the application will decide which one of these will suit the application best.

Buoyed by the exciting application possibilities of the blockchain technology, hundreds of startups have sprung up all over the world. These startups are building innovative solutions in a wide variety of domains. Startups like Consensys, Storj, Ripple, NEO, Bancor are creating solutions targeted at different industries. In India too, there are dozens of startups which have ventured into creating smart solutions based on blockchains. 

\subsubsection{Applications in the Financial Sector}
Financial institutions in general and banking and insurance sectors in particular have started embracing blockchain technology because of a variety of robust features that blockchain technology provides: immutability, security, trustworthiness, and intelligent execution of smart contracts. As one example, more than 80 of the world's major financial institutions are set to use a distributed ledger platform {\bf CORDA} which is a blockchain based system developed by a prominent financial technology startup {\bf r3}. The insurance sector is also pursuing the blockchain technology aggressively. See \cite{DELOITEE2017}, \cite{RBI2017},  \cite{SPECTRUM2017},  and  \cite{UNDERWOOD2016} for more details. 

\subsubsection{Applications in the Supply Chain and Logistics Sector}
We foresee the supply chain and logistics sector to have the highest promise for a rich variety of usecases for application of blockchain technology. Usecases such as tracking food items from agricultural farms to customer delivery, tracking the numerous parts of a car that will circulate through various suppliers, manufacturers, and countries, and identifying contamination and reducing waste in perishable materials and food processing; etc. represent examples of situations where blockchain technology is valuable. A startup called everledger provides a global digital ledger that tracks and protects valuable assets throughout their lifetime. The digital incarnation on blockchains is used by various stakeholders across the supply chain to form provenance and verify authenticity. This has been used effectively for example in tracking diamonds; more than 1.6 million diamonds are reportedly captured in the blockchain created.

We believe an intriguing  application of blockchains in the supply chain context is in {\em dynamic supply chain formation\/}. Often, companies need to put together a supply chain of partners dynamically to respond to dynamically arriving demands. This may throw up the need to partner with unknown entities. The trust provided by the blockchain framework provides a brilliant opportunity for partners who are unknown to one another to come together and dynamically form a supply chain for the specific immediate purpose.

\subsubsection{IoT Based Applications}
In verticals such as healthcare, smart cities, intelligent buildings, home automation, energy industry, supply chains, and manufacturing, the application of IoT technology is well known and well documented. In these applications, typically, hundreds and often thousands of smart devices are connected to an Internet server or cloud, and, a variety of analytics based services are provided. These services, however, currently suffer from lack of scalability, challenges of security, lack of standards, high costs, and moreover are centralized. Blockchain technology can be harnessed to provide a distributed, scalable  solution where smart devices now connect directly to distributed ledgers. Moreover, security guarantees can be provided by robust blockchain platforms. See for example \cite{IBM2017} for more details.

\subsubsection{Block Chains for Social Good}
There is a wide repertoire of applications which are citizen centric and social welfare oriented where blockchains hold significant promise. These are discussed in detail in \cite{DELOITEE2017},  \cite{IANSITI2017}, \cite{SPECTRUM2017}, \cite{TAPSCOTT2016}, and  \cite{UNDERWOOD2016}.

\noindent In the education sector, safe deposit of academic transcripts and educational records is a usecase that is already being pursued by many startups and educational institutions. In the healthcare area, management of electronic medical records to ensure privacy guarantees and to enable analytics-based diagnosis, etc, are prominent examples where blockchain technology is being explored. In context of energy systems, monitoring smart grids using a blockchain based, IoT based framework is already popular. 

Blockchain technology has excellent potential for promoting ethical trade. For example, a startup Mediledger provides blockchain solutions for pharmaceutical supply chains. In one usecase, the blockchain systematically tracks pharma drugs from manufacture to ultimate delivery by making drug companies, hospitals, distributors, and wholesalers to record drug deliveries on the blockchain. This will enable to verify the authenticity and provenance of the drugs, thus leading to elimination of counterfeit drugs.

\section{Conclusion}
In this chapter, we have discussed key technical ingredients of the blockchain technology and made a brief survey of its tremendous application potential. Substantial investments have been made world over by corporations and Governments on implementing blockchain technology in multiple sectors and a large number of startups have sprung up in the past few years to harvest this technology in highly innovative ways.

There are at least two major reasons why blockchain technology will survive the test of time. The first reason is it is deep technology founded on solid fundamental principles and technical rigor. Such a technology will always blossom, perhaps, in a slightly different form. The second reason is blockchain technology has the promise and potential to touch billions of people. It has the potential to enable vast improvements to existing  applications  and help launch radically new applications. One such application is enabling privacy-preserving B2B collaboration.


\chapter{Problem Formulation and Approach}
\begin{quote}
{\em This chapter is devoted to formulating the requirements of the B2B platform and providing a formal framework to capture, study, and analyze the interactions of agents on such a platform. Further the model of the agents is presented, which forms the basis of the analysis presented in the following chapter. The chapter concludes with a short description of the cryptographic protocols which help fulfill the requirements of the B2B platform, and validate the assumptions made in the analysis detailed in the following chapter.}
\end{quote}
\section{Requirements of the B2B Platform }
\label{sec:srp}

\subsection*{Motivation for B2B Collaboration}
Online B2B platforms are almost as old as the Internet itself.  Electronic Data Exchange (EDI) brought the first wave of B2B transformation, providing significant benefits, especially, in the automotive and retail sectors \cite{mukhopadhyay2002strategic}. However, the adoption of EDI and related mechanisms has not progressed beyond point-to-point messages and hence, not enabled ecosystem collaborations like supplier identification/selection, price negotiation, collective quality improvement, etc \cite{tanner2008current}. Apart from cost of EDI, another reason for the stalled progress has been the lack of a framework for collaboration which protects business sensitive information. 

B2B collaboration may appear to be rather paradoxical. Expecting businesses, who may be competitors, to cooperate, sounds implausible, however, there have been several studies that encourage it. Collaboration amongst businesses has been shown to be helpful in reducing supply chain costs both theoretically as well as empirically \cite{cachon2000supply, doboscooperation, ganesh2014distribution}. Furthermore, many studies show that to ensure continued quality in the products purchased from a strategic agent, it is necessary for all other buyers to know when a particular seller fails to deliver good quality \cite{fainmesser2012community,gazzale2011remain}. 

\subsection*{Roadblocks}
A typical online market platform, such as Amazon, Flipkart, eBay et cetera, allows for this by encouraging buyers to give feedback about the products they receive. Customers on these platforms accordingly make their decision based on the price quoted by the seller and the quality perceived from their own personal experience and the feedback regarding that seller. Enterprise agents would also like to have such information at hand to be able to choose the right supply chain partners \cite{haghpanah2011trust}. However, giving such feedback is not a straightforward matter for enterprise buyers and sellers. Enterprise agents cannot have their identity and buying history be leaked by way of such feedback, and anonymous feedback in itself is not completely reliable. As a result, while there is a pressing need for a Business-to-Business collaboration platform, the sensitive nature of the information that must be exchanged on such a platform restricts the actual development.  

\subsection*{Components of a B2B platform}
On any B2B exchange, it is common to have a group of {\em competing manufacturers} (buyers) using the platform to interact with a common base of \emph{suppliers} (sellers) to automate their procurement process. While they compete at an overall business level, they would like to cooperate for purposes of rating the suppliers for their quality, timeliness, compliance, et cetera, for collectively eliminating systemic inefficiencies. However, this is not possible at present because the available platforms do not have any functionality that allows collaboration without revealing business sensitive information such as who is transacting with whom. 

Permissioned blockchain networks bring the necessary levels of trust and data security along with the benefits of automation through smart contracts. There is a dire need for combining blockchain technology's strengths with new ideas to ensure that agents can collaborate without revealing business sensitive information. In the context of rating suppliers for their quality, this amounts to the following: the buyers are able to collaborate in a way that they compute credible quality rating for the sellers with the guarantee that the buyers are incentivized to be honest and without revealing who buys from whom or what rating they have given. We achieve precisely this in our work with a confluence of ideas from game theory, encrypted private computations, and blockchain technology.

In this thesis, we are motivated to undertake a principled study of how the blockchain technology facilitates better B2B collaborations across multiple networks. We propose a novel framework for orchestrating trusted and privacy-preserving B2B collaboration on blockchain platform. Further, we consider a prototypical example of B2B collaboration for implementing supplier rating system. We show how techniques from game theory and cryptography can be used for implementing such a system.  

We formulate a simple, stylized model that captures key aspects of a collaborative procurement process using a repeated game formulation. The motivation for choosing such a model is that it helps to bring out the key aspects of the problem without any diversions of actual business process. We consider an online B2B platform on which enterprise buyers and sellers interact repeatedly. Each seller can deliver either high quality goods or low quality goods. It costs a seller $c$ to produce high quality and $0$ to produce low quality. Further, the goods may be sold at a price fixed by the platform or different sellers may adopt different pricing strategies. The procurement happens in multiple rounds and in each round, a buyer chooses the seller to buy from. Further, each buyer gives a binary feedback on the quality of the received supply. 


If an oracle (a trusted third party with access to all the feedback), had to publish ratings for the sellers, these ratings would simply be the weighted average of the feedback that a seller gets from different buyers across all the rounds, with more importance for recent rounds. However, the assumption of a trusted third party itself is implausible, even more so for competing enterprise agents. Consequently, we look to blockchain technology to provide a decentralized solution.

The problem is to first design a rating system which aggregates the buyer feedback such that:
\begin{itemize}
\item Only the transacting parties get to know who bought from whom in any of the rounds, that is, it does not become public knowledge which buyer bought from which seller in any given round; 
\item Sellers cannot decipher the exact feedback given about by a particular buyer in a specific round, unless everybody has given only low feedback or only high feedback;
\item The mechanism disincentivizes the buyers from providing dishonest feedback by means of a smart contract in place to penalize buyers suitably if feedback is found to be dishonest.
\end{itemize}
The design of  such a {\em strategy proof market platform\/} (or simply {\em strategy proof mechanism\/}) involves many challenges, which we solve in this thesis, by drawing upon game theoretic analysis and smart contracts deployed on permissioned blockchains and multi-party secure computation. To achieve the aforementioned objectives, we propose a permissioned blockchain that implements an appropriate smart contract, in conjunction with regulation protocols (a public perception protocol and a monitoring protocol).  
Given such a marketplace that has a blockchain infrastructure and regulation protocols in place, we next explore different pricing rules and punishment strategies in a game theoretic setting with the objective of identifying mechanisms that maximize social welfare (that is, induce high quality goods to be produced most often).

\section{A Framework for a Blockchain based B2B Platform} 

Having listed the requirements from the B2B platform, we introduce a formal framework to model and analyze events that take place on the B2B platform, and thus on the permissioned blockchain. To the best of our knowledge, this is the first notation used to formally capture the events on a blockchain-based platform. One of the purposes of this notation is to help capture and study the information shared on the platform, and whether any of the constraints discussed in \ref{sec:srp} are violated. Another motivation for this framework is to formally encapsulate the nature of the blockchain ledger as a set of sequential events. An event on the blockchain may be a consequence of any subset of prior events. The framework assumes that the blockchain is a permissioned one. We use a permissioned blockchain setup as it helps provide a simple to implement reputation system and additionally provides provenance to all the events and information. The framework also provides notations for the analysis protocols deployed on such permissioned blockchains with regards to the anonymity properties satisfied. \\

\noindent Let $\mathcal{N}$ be the set of participants (aka nodes) in a permissioned blockchain network. \\

\noindent Let $\mathcal{E}$  be the set of all transactions (aka events) that occur as a part of collaboration between nodes in $\mathcal{N}$. 

For each event $e \in \mathcal{E}$, there is an associated set of participants represented by $\mathcal{N}_e \subseteq \mathcal{N}$; it represents the set of participants involved with the event $e$. 

Further, for each event $e \in \mathcal{E}$, there is an associated record denoted by $\mathcal{R}_e$; this captures the outcome of the event $e$. Participants  in the network may have selective access to the information about an event. \\

\noindent Let $D_{h, \mathcal{R}_e}$  denote the subset of data fields of $R_e$ that can be accessed by  the node $h \in N_e$.  \\

\noindent Thus, $D_h = \cup_{e \in \mathcal{E}} D_{h, \mathcal{R}_e}$  is the overall data that is accessible to participant $h$. \\

\noindent We denote the union of all accessible information in the blockchain network by $D_U$, i.e, $D_U = \cup_{h \in \mathcal{N}} D_h$. 

Consider a function $\Gamma(\mathcal{N}, \mathcal{E}, D_{U})$ that can be computed if there is an oracle with access to the information $D_U$. \\

\noindent Consider a protocol $\mathcal{P}$ which  
\begin{enumerate}
\item Consists of an \textbf{ordered set} of  encrypted messages $\mathcal{S}$ generated by the nodes in $\mathcal{N}$, and, 
\item Executes a sequence of computation steps, denoted by $\mathcal{C}$  with each step being carried out by a subset of nodes, where, each step may consume any subsequence of previously generated messages. 
\end{enumerate}
Suppose the protocol computes a function $\Gamma^{'} (\mathcal{S}, \mathcal{C})$. Conceptually, one may regard the last step as a step carried out by a set of special {\em monitor} nodes.

The protocol $\mathcal{P}$  is said to be a \textbf{secure and private simulation} of $\Gamma(\mathcal{N}, \mathcal{E}, D_{U})$ if it satisfies the following conditions:
\begin{enumerate}
\item $\Gamma(\mathcal{N}, \mathcal{E}, D_{U})$  $\approx$ $\Gamma^{'} (\mathcal{S}, \mathcal{C})$
\item If $h \notin \mathcal{N}_e$, then, even with access to $\mathcal{S}$ and $\mathcal{C}$ , $h$ cannot infer the exact set $N_e$
\item No node $h \in \mathcal{N}$ can infer any data belonging to $D_{U} \setminus D_h$
\end{enumerate}
We regard, $\mathcal{S}$, the set of messages generated by all the nodes, as constituting the blockchain ledger. A B2B collaboration is essentially a sequence of computation steps $[\Gamma_i, \forall i = 1, 2, \ldots]$ We say that the collaboration can be securely and privately simulated on the blockchain if each of the steps has a secure and private simulation protocol. This framework is general enough to capture a host of B2B collaboration scenarios.\\

\noindent For our current problem, $\mathcal{N}$ is the set of all buyers and sellers. \\

\noindent Each transaction between a buyer-seller pair is represented by an event $e \in \mathcal{E}$. \\

\noindent Associated with each event $e$ is the record $<$price, cost, quality, rating$>$;  the buyer knows $<$price, quality, rating$>$ while the seller knows $<$price, cost, quality$>$. \\

\noindent Our goal is to compute the supplier rating that closely approximates the rating that an oracle with complete access would compute while satisfying trust and privacy issues captured in Section \ref{sec:srp}.


\section{ Building Blocks of the Framework }
\subsection{The Model}
We now set up the model of the procurement process to be studied.

Let $B$ be the set of buyers and $S$, the set of sellers. Therefore, $\mathcal{N}= B \cup S$

Let $n_B$ and $n_S$ be the number of buyers and number of sellers, respectively. We assume that $n_S>2$ and $n_B>2$; otherwise, irrespective of the system, anonymity constraints can never be met. If $n_S <3$, then each seller knows that the buyer who did not buy from him, chose to buy from the other seller, if any. If $n_B < 3$, the sellers who have made a sale know which buyer has bought from whom, as if a buyer has not bought from him, she must have bought from the only other seller who has made a sale. We assume that the list of sellers who have made at least one sale in a given round is made public at the end of that round. Thus, it is imperative to assume that both $n_B$ and $n_S$ are strictly greater than 2.

The recurring interactions are modeled as a repeated game. We formally define repeated games in the next chapter. Each round of the repeated game takes place as follows:
\begin{enumerate}
\item Each buyer selects a seller to buy from and places an order
\item Each seller decides what quality to give to which buyer
\item Based on the quality received, the buyer submits encrypted feedback
\end{enumerate}

\begin{figure}[h!]
  \centering
  \begin{subfigure}[b]{0.3\linewidth}
    \includegraphics[width=\linewidth]{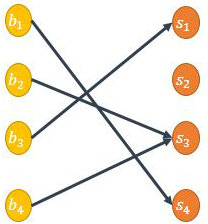}
    \caption{Buyers choose seller}
  \end{subfigure}
   \begin{subfigure}[b]{0.3\linewidth}
    \includegraphics[width=\linewidth]{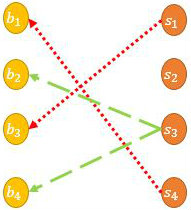}
    \caption{Sellers decide the quality}
  \end{subfigure}
   \begin{subfigure}[b]{0.3\linewidth}
    \includegraphics[width=\linewidth]{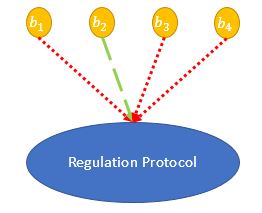}
    \caption{Buyers submit feedback}
  \end{subfigure}
  \caption{An example round.\\ {\small Red-dotted arrows and green-dashed arrows denote low and high quality respectively}}
\end{figure}
\noindent The set of sellers who are making at least one sale is shared with all buyers and sellers through a smart contract at the end of each round. The buyers then deploy the public perception protocol, amongst themselves, for calculating the ratings of the sellers.

Each buyer $b$ maintains a personal history vector $h_b^t$ which denotes the discounted average number of times the sellers have given buyer $b$ a high-quality product. For those sellers, with whom the buyer has never transacted, $\xi$ is the initial perception. $\xi$ is also the rating for sellers who are yet to make a sale. Thus, $Q^1 (s)=\xi$, $\forall s$. 
\begin{equation}
h_b^t(s)= \frac{\sum _{i=1}^{t-1} \delta _b ^{t-i-1}I_Q(s,b,i) +\bar{\xi}(s)\Pi _{i=1}^{t-1}(1-I_T(s,b,i))}{\sum _{i=1}^{t-1} \delta _b ^{t-i-1}I_T(s,b,i) + \Pi _{i=1}^{t-1}(1-I_T(s,b,i))} 
\end{equation}
where $I_Q(s,b,i)$ is the indicator function of whether seller $s$ gave buyer $b$ a high-quality product in round $i$, and $I_T(s,b,i)$ is the indicator function whether seller $s$ sold  to buyer $b$ in round $i$. 

$\bar{\xi}(s)$ is $\xi$ in round $t$ if $\xi \geq Q^i(s),\, \forall i=1,\cdots, t$ else is $Q^t(s)$. 

We model the buyer as being somewhat optimistic in believing the initial perception if the public perception is not as good, but once a buyer starts believing the public perception, she does not revert to the initial perception. 

The discounting factor is $\delta _b$ $\in (\tau,1)$, $\tau>0.5$. This value is private information of the buyer, and is not revealed during the game. 

At the start of each round, a buyer $b$ will compute the likelihood perceived by her that seller $s$ will give a high quality product in that round. She does so by taking a weighted average of her own history with the seller, and the seller's public perception. This gives the buyer's personal perception vector: 

\begin{equation}
q_b ^t (s)= \theta _b h_b ^t (s) + (1-\theta _b)Q^t (s)
\end{equation}
where $\theta _b$ is the type of the buyer, assigned by nature. $\theta _b$ denotes the weight that buyer $b$ places on her own history with a seller versus the seller's public perception. 

\noindent Expected utility of a buyer from a seller $s$ is \[ q_b ^t(s)v_H + (1- q_b ^t (s))v_L - p(s)\] where $p(s)$ is the price charged by seller $s$. Each buyer will choose the seller who maximizes her expected utility. If there are multiple such sellers, then the buyer chooses the one who most recently gave a high-quality product if any, else chooses randomly among them. Therefore, the first choice of seller will always be random.

For a seller, as all buyers are equivalent, we assume that a seller will always follow the same strategy for all buyers. 

Seller $s$ has a discounting factor $\sigma _s$ for future payoffs, which determines the strategy chosen. Thus, choosing a payoff of $p$ has utility $p$ in the current round, $\sigma_sp$ in the next round, $\sigma _s ^2p$ in the round after that, and so on. We elaborate on this when we cover repeated games in the next chapter. 

\subsection{Cryptographic Protocols}
Based on the model defined above, we propose a Multi-Party Computation (MPC) protocol to compute the public perception vector and to ensure that the requirements listed in \ref{sec:srp} are satisfied. In order to do so, we utilize the blockchain structure and make an additional assumption that the indicator function $I_T(s,i)$, which denotes whether or not a seller $s$ has made at least one sale in round $i$, is public knowledge to all entities. This can easily be ensured by means of a smart contract that is deployed across all the channels on the platform. We call the aforementioned MPC protocol as the Regulation protocol. This protocol is deployed amongst the buyers at the end of each round, by means of a smart contract.

The core structure of the regulation protocol is as follows:
\begin{itemize}
\item The protocol contains a set of monitors who together with a software entity (SE) are responsible for identifying dishonest feedback monitor the behavior of buyers. 
\item Each buyer shares her encrypted feedback among the buyers and they jointly compute the components of public perception vector using homomorphic encryption.
\item Finally, the buyers utilize the results of the computation to obtain $Q^t(s)$.
\end{itemize}
This protocol is essential to the platform as it provides a secure method for computing ratings for sellers, without revealing buying histories, both of which are critical for buyers. Further, it validates the assumptions made in the analysis presented in the next chapter. The details of this protocol are presented in Chapter 5. 

\subsubsection{The Rating System}
We now define the exact formulation of the ratings that are computed by the regulation protocol. As mentioned earlier, if an oracle were to compute these ratings, she would simply consider the  fraction of sales made by the seller which are rated high, giving more weight to recent rounds. We attempt to replicate these ratings in a decentralized fashion. 

The  public perception of a seller $s$, calculated by the regulation protocol, is the discounted average of the fraction of sales made by $s$ which were given a high feedback, using random parameters. The data taken from each round is whether a seller has made a sale in that round, and if so, what fraction of the sales were reported to be of high-quality. The data from round $i$, in the current round $t$ has weight $\delta ^{t-i-1} _M$. Thus, the most recent round has weight 1, the one before that has weight $\delta _M$ and so on. Rounds in which no sales were made have weight 0. The weighted average of this data is the rating of the seller.


\noindent The public perception vector is computed in round $t$ as follows:

\begin{equation}
 Q^t (s)= \frac{\sum _{i=1}^{t-1} \delta _M ^{t-i-1}I_Q(s,i) + \xi \Pi _{i=1}^{t-1}(1-I_T(s,i))}{\sum _{i=1}^{t-1} \delta _M ^{t-i-1}I_T(s,i) + \Pi _{i=1}^{t-1}(1-I_T(s,i))}
\end{equation}
where $I_Q(s,i)$ is the fraction of sales made by seller $s$ in round $i$ that got a feedback of being high-quality, and $I_T(s,i)$ is the indicator function whether seller $s$ made any sales in round $i$.
If seller $s$ has not made any sale till round $t$ then $Q^t (s)=\xi$.
The value of $\delta _M$ is chosen uniformly at random from the interval $(\bar{\tau},1)$ in each round. This interval is agreed upon by all the buyers, prior to the commencement of trading on the market. This is done to prevent sellers from deciphering the exact feedback given for them. As the value of $\delta _M$ used for all sellers is the same, there is no partisanship. 

Henceforth, we use the terms `public perception' of a seller and his `rating' interchangeably.

\section{Conclusion}
A platform where enterprise buyers jointly compute the rating or reputation of the sellers is not without constraints. The aggregation of the feedback must preserve the anonymity of the feedback given by each agent, at the same time incentivize these agents to be honest in their feedback. The details of the monitoring protocol which will be deployed are covered in Chapter 5. The interactions between buyers and sellers are further modeled as a repeated game. The outcome of these interactions under various pricing rules is studied in the following chapter.

\chapter{Game Theory Based Smart Contracts}
\begin{quote}
{\em In this chapter we present the analysis of equilibrium behaviour of agents on a B2B market platform where feedback from enterprise buyers is aggregated in a privacy-preserving manner. We first cover necessary game theory preliminaries: strategic form games in Section 4.1, extensive form games in Section 4.2, repeated games and what punishment means in the context of the supplier rating problem in Section 4.3. The full analysis follows these preliminaries. In Section 4.4, we study the case where  sellers sell at the same price, and within this case study different punishment strategies, in order to identify which strategy best incentivizes sellers to produce high-quality. We analyze two other pricing rules, namely binary and continuous pricing, in Sections 4.5 and 4.6 respectively. All these cases follow the model discussed in the previous chapter, where the procurement process is modeled as a repeated game.}
\end{quote}

\section{ Strategic  Form Games and Nash Equilibrium}
Game theory is a study of interactions between strategic agents that are rational and selfish. Agents or players know the game and act to maximize their own payoff or utility. We first define Strategic Form Games and Nash Equilibria. Definitions in this section are taken from \cite{narahari2014game}.

\begin{definition} A strategic form game $\Gamma$ is a tuple $<N, (S_i)_{i\in N}, (u_i)_{i\in N}>$ where

\begin{itemize}
\item $N= \{1,2,\cdots,\,n\}$ is a finite set of players
\item $S_1, S_2 ,\cdots ,\, S_n$ are the strategy sets of the players $1,2,\cdots,\,n$
\item $u_i:S_1 \times S_2 \times \cdots \times S_n \rightarrow \mathbb{R}$ for $i= 1,2, \cdots, n$ are mappings called the utility functions or payoff functions
\end{itemize}
\end{definition}
\noindent Strategic form games assume that all players act simultaneously, with no knowledge of what strategy other players are going to choose.
We give the example of a well known strategic form game called Prisoner's Dilemma in Table 4.1. The game captures a setting where two prisoners, Anita and Bharat, are suspected of having committed a crime and are being interrogated in separate rooms. Each of them has the option of confessing or not confessing. If neither confesses, they both get minimal punishment of $-1$ each. If only one confesses, then that person does not get punished, and the other gets the entire punishment for the crime $-10$. If both confess, the punishment is divided amongst the two and each get $-5$.

\begin{table}[h]
\renewcommand\arraystretch{1.5} 
\centering
\begin{tabular}{ll | c | c |}
&\multicolumn{1}{c}{}&\multicolumn{2}{c}{\textbf{Anita}}\\[-2ex]
&\multicolumn{1}{c}{}
&\multicolumn{1}{c}{Confess}&\multicolumn{1}{c}{Not Confess}\\
\cline{3-4}
\multirow{2}{*}{\rotatebox{90}{\textbf{Bharat}}}
&Confess&-5,-5&0,-10\\
\cline{3-4}
&Not Confess&-10,0&-1,-1\\
\cline{3-4}
\end{tabular}
\caption{The Prisoner's Dilemma}
\label{table:pd}
\end{table}
\noindent Clearly, the socially optimal outcome would be when neither of them confess. However, it is easy to check that whether Anita confesses or not, Bharat will be better off confessing and vice versa. Thus, the outcome of this game, when played by rational and selfish agents, is having both of them confess. This is because Anita's best response to Bharat confessing is to confess herself. A strategy profile where the strategy chosen by each player is also their best response to all the other players' chosen strategies is called a Nash Equilibrium.

\begin{definition}[Pure Strategy Nash Equilibrium] A strategy profile $s^*\,=\, (s^*_1,\cdots,\, s^*_n)$ is said to be a pure strategy Nash equilibrium of the game  $\Gamma = <N, (S_i)_{i\in N}, (u_i)_{i\in N}>$ if \[u_i(s^*_i,s^*_{-i})\geq u_i(s_i,s^*_{-i}),\,\, \forall s_i\in S_i, i\in N \]
\end{definition}
\noindent Thus in the Prisoner's Dilemma game, $(\textit{Confess,Confess})$ is a Pure Strategy Nash Equilibrium (PSNE). A strategic form game need not have exactly one PSNE. Given below are two different strategic form games, one with two PSNE and one with none. 

\begin{table}[h]
\renewcommand\arraystretch{1.5} 
\centering
\begin{tabular}{ll | c | c |}
&\multicolumn{1}{c}{}&\multicolumn{2}{c}{\textbf{Dhruv}}\\[-2ex]
&\multicolumn{1}{c}{}
&\multicolumn{1}{c}{B}&\multicolumn{1}{c}{S}\\
\cline{3-4}
\multirow{2}{*}{\rotatebox{90}{\textbf{Chitra}}}
&B&2,1&0,0\\
\cline{3-4}
&S&0,0&1,2\\
\cline{3-4}
\end{tabular}
\caption{Bach or Stravinsky}
\label{table:bos}
\end{table}
\noindent This game, called Bach or Stravinsky, captures a setting where two friends need to simultaneously  decide whether to attend a performance by Bach or one by Stravinsky without consulting the other person. While Chitra prefers Bach to Stravinsky, Dhruv prefers Stravinsky to Bach, but neither enjoy going alone. As a result, if Chitra were to know where Dhruv is going, irrespective of which of the two strategies David chooses, she would do the same, and vice versa. Consequently, this game has two PSNE, namely $(B,B)$ and $(S,S)$. 

We now study a commonly played game called Matching Pennies. Two players, Vandana and Kishor choose a side each. If both choose the same side, Alex wins, otherwise Sara wins. This game is denoted as a strategic form game in Table 4.3.

\begin{table}[h]
\renewcommand\arraystretch{1.5} 
\centering
\begin{tabular}{ll | c | c |}
&\multicolumn{1}{c}{}&\multicolumn{2}{c}{\textbf{Vandana}}\\[-2ex]
&\multicolumn{1}{c}{}
&\multicolumn{1}{c}{H}&\multicolumn{1}{c}{T}\\
\cline{3-4}
\multirow{2}{*}{\rotatebox{90}{\textbf{Kishor}}}
&H&1,-1&-1,1\\
\cline{3-4}
&T&-1,1&-1,-1\\
\cline{3-4}
\end{tabular}
\caption{Matching Pennies}
\label{table:mp}
\end{table}
\noindent It is easy to see that this game does not have a PSNE. If both choose the same side of the coin, Vandana will want to change her choice. If they don't Kishor will want to change. However, if Kishor and Vandana were to choose a probability distribution with which to choose a side, and then leave the actual choice to the toss of a private coin, the choice of probability distribution would indicate a {\em Mixed Strategy}.

\begin{definition}[Mixed Strategy] Given a player i with the set of pure strategies $S_i$, a mixed strategy $\sigma _i$ of player $i$ is a probability distribution over $S_i$.  That is, $\sigma _i: S_i \rightarrow [0,1]$ assigns to each pure strategy $s_i\in S_i$, a probability $\sigma _i(s_i)$ such that $\sum _{s_i \in S_i} \sigma _i(s_i)=1$.
\end{definition}
\noindent Each pure strategy of a player is also a degenerate mixed strategy. The set of all probability distributions on $S_i$ is denoted as $\Delta (S_i)$, and is often called the {\em mixed extension} of $S_i$. Thus,

\[\Delta (S_i)=\{\sigma \in \mathbb{R}^{|S_i|}_+: \sigma _i(s_i)=1\}\]
A Nash Equilibrium can be defined using mixed strategies as follows:
\begin{definition}[Mixed Strategy Nash Equilibrium] A mixed strategy profile $\sigma^*\,=\, (\sigma^*_1,\cdots,\, \sigma^*_n)$ is said to be a mixed strategy Nash Equilibrium of the strategic form game  $\Gamma = <N, (S_i)_{i\in N}, (u_i)_{i\in N}>$, if \[u_i(\sigma^*_i,\sigma^*_{-i})\geq u_i(\sigma_i,\sigma^*_{-i}),\,\, \forall \sigma_i\in \Delta(S_i), i\in N \]
\end{definition}
\noindent All of the aforementioned games have Mixed Strategy Nash Equilibria (MSNE). For the Matching Pennies game, $\sigma _i(s_i)=\frac{1}{2}, \forall s_i\in S_i, \forall i\in N$ is the MSNE. In fact, Nash has proven that every finite strategic form game has a MSNE.

\begin{theorem}[Nash \cite{nash1951non}] \label{thm:nash}
Every finite strategic form game $\Gamma = <N, (S_i)_{i\in N}, (u_i)_{i\in N}>$ has at least one mixed strategy Nash equilibrium
\end{theorem}
\section{Extensive Form Games and Subgame Perfect Equilibrium} 

Strategic form games assume that players move simultaneously, with no knowledge of the other players' choices. However, a large class of games, such as chess, do not belong to this paradigm and are actually played sequentially, and players may or may not be able to observe the choice made by other players when it is their turn play their own strategy. Extensive form games provide a framework to capture such games. The definitions in this section have been taken from \cite{narahari2014game}.

\begin{definition}[Extensive Form Games] An extensive form game $\Gamma$ consists of a tuple $\Gamma = <N, (A_i)_{i\in N}, \mathbb{H}, P, (\mathbb{I}_i)_{i\in N}, (u_i)_{i\in N}>$ where

\begin{itemize}
\item $N= \{1,2,\cdots,\,n\}$ is a finite set of players.
\item $A_i$ for $i=1,2,\cdots , n$ is the set of action available to player $i$ (action set of $i$).
\item $\mathbb{H}$ is the set of all terminal histories where a terminal history is a path of actions from the root to a terminal node such that it is not a proper subhistory (including the empty history $\varepsilon$) of all terminal histories.
\item $P: S_{\mathbb{H}} \rightarrow N$ is a player function that associates each player subhistory to a certain player.
\item $\mathbb{I}_i$ for $i=1,2,\cdots , n$ is the set of all information sets of player $i$.
\item $u_i:\mathbb{H}\rightarrow \mathbb{R}$ for $i=1,2,\cdots , n$ gives the utility of player $i$ corresponding to each terminal history.
\end{itemize}
\end{definition}
\noindent Choosing a strategy in an extensive form game doesn't seem the same as choosing one in a strategic form game. At different stages in the game, different actions may be possible. Given an information set $J\in \mathbb{I}_i$, let $C(J) \subseteq A_i$ be the set of all actions possible to player $i$ in the information set $J$. A strategy is formally defined as follows

\begin{definition}[Strategy]
A strategy $s_i$ of player $i$ is a mapping $s_i: \mathbb{I}_i \rightarrow A_i$ such that $s_i(J)\in C(J)\forall J \in \mathbb{I}_i$.
\end{definition}
\noindent Thus, a strategy maps every possible information set to an action. The strategies chosen by all the players in the game determine the outcome of the game.

\begin{definition}[Outcome]
Given an extensive form game $\Gamma$ and a strategy profile $s=(s_1, \cdots, \, s_n)$ in the game, the outcome resulting from the terminal history corresponding to the strategy profile $s$ is called the outcome of $s$ and is denoted by $O(s)$. 
\end{definition}
\noindent An example of an extensive form game is given in Figure 4.1, called the Entry game. A challenger must decide whether or not to enter a new industry and accordingly challenge the incumbent, who's already present in the industry. If the challenger does not choose to enter, then the challenger and incumbent get a utility of 1 and 2 respectively. If the challenger does enter, the incumbent must choose to either fight her or accommodate her. If he chooses to fight her then both get a payoff of $0$ else, the challenger gets a payoff of $2$ and the incumbent a payoff of $1$. 

\begin{figure}[h!]
\label{fig:entrygame}
\hspace{0.25\linewidth}
\includegraphics[width=0.5\linewidth]{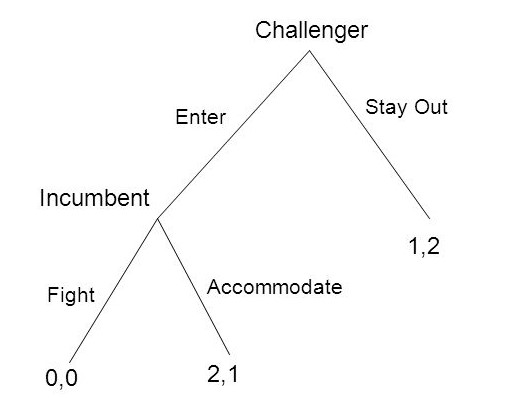}
\caption{Entry Game}
\end{figure}

\begin{definition}[Subgame] Given an extensive form game $\Gamma$ and a non-terminal history $h$, the subgame following $h$ is the part of the game that remains after the history $h$ has occurred.
\end{definition}
\noindent A pure strategy Nash equilibrium for extensive form games is defined as follows:

\begin{definition}(Pure Strategy Nash Equilibrium (PSNE)) 
Given an extensive form game $\Gamma =<N, (A_i)_{i\in N}, \mathbb{H}, P, (\mathbb{I}_i)_{i\in N}, (u_i)_{i\in N}>$, a strategy profile $s^*=(s^*_1, \cdots, \, s^*_n)$  is called a pure strategy Nash Equilibrium if $\forall i \in N$, \[u_i(O(s^*_i,s^*_{-i}))\geq u_i(O(s_i,s^*_{-i})),\,\, \forall s_i\in S_i\]
\end{definition}
\noindent Essentially, a pure strategy Nash equilibrium checks that each player's chosen strategy is optimal given the strategy chosen by other players, at the start of the game. It only takes into account the terminal history that actually occurs as a result of the strategy profile, and does not consider other subgames. If we were to do so, we would be considering a different notion of equilibrium called Subgame Perfect Equilibrium (SGPE).

\begin{definition}(Subgame Perfect Equilibrium)
Given an extensive form game $\Gamma =<N, (A_i)_{i\in N}, \mathbb{H}, P, (\mathbb{I}_i)_{i\in N}, (u_i)_{i\in N}>$, a strategy profile $s^*=(s^*_1, \cdots, \, s^*_n)$  is an SGPE if $\forall i \in N$, \[u_i(O_h(s^*_i,s^*_{-i}))\geq u_i(O_h(s_i,s^*_{-i})),\,\, \forall h \in \{x\in s_{\mathbb{H}}: P(x)=i \}\forall s_i\in S_i\]
where $O_h(s^*_i,s^*_{-i})$ denotes the outcome corresponding to the history $h$ in the strategy profile $(s^*_i,s^*_{-i})$.
\end{definition}
\noindent Clearly, a PSNE is also a SGPE. The PSNE of the Entry Game are (Stay Out, Fight) and (Enter, Accommodate). However, the only SGPE of this game is (Enter, Accommodate).

\section{Repeated Games} 
As seen earlier, in the Prisoner's Dilemma game, as the agents try to maximize their own payoff, they end up reaching an outcome that is not pareto optimal. As there is nothing to deter them from making the selfish (and rational) choice of confessing, they choose to do so. A strategic form game does not capture the consequences of the decision and does not facilitate a "threat of punishment". To do so we study repeated games. The definitions in this section have been adapted from \cite{osborne2004introduction}.
\begin{definition} [Repeated Game]
Let $\Gamma = <N, (S_i)_{i\in N}, (u_i)_{i\in N}>$ be a strategic form game. The $T$-\textbf{period repeated game} of G for the discounting factor $\delta$ is the extensive game with perfect information and simultaneous moves where

\begin{itemize}
\item $N$ is the set of players.
\item the set of terminal histories is the set of sequences $(s^1,s^2,\cdots,\, s^T)$ of action profiles in $\Gamma$.
\item $P$ assigns the set of all players to every history $(s^1,s^2,\cdots,\, s^t)$ for every value $t$.
\item the set of actions available to player $i$ after any history is $S_i$.
\item each player $i$ evaluates each terminal history $(s^1,s^2,\cdots,\, s^T)$ according to the discounted sum $\sum _{t=1}^T \delta ^{t-1}u_i(a^t)$.
\end{itemize}
\end{definition}
\noindent The \textbf{infinitely repeated game of } $\Gamma$ for the discount factor $\delta <1$ differs only in that the set of terminal histories is the set of infinite sequences $(s^1,s^2,\cdots )$ and the utility to each player $i$ to the terminal history $(s^1,s^2,\cdots )$ is the discounted sum $\sum _{t=1}^\infty \delta ^{t-1}u_i(a^t)$.

In a repeated game, if a player does not cooperate in one round, they may have to face some punishment by the other players in future rounds. In the example of Prisoner's Dilemma, it is beneficial for both the players, if neither of them confess. Irrespective of what strategy a player chooses, they will be worse of if the other player decides to confess. Thus, the players can mutually agree not to confess and that if one of the players confesses in a round, then the other will confess as punishment in future rounds. 

Typically, three types of punishment strategies in repeated games are studied:

\begin{itemize}
\item Grim Trigger: Under grim trigger strategies, if one player does not cooperate in one round, then all other player will punish the defecting player in all future rounds.
\item Tit for Tat: Under tit for tat, if a player does not cooperate in a round, then she will be punished in the next round.
\item Limited Punishment: Under limited punishment, if a player does not cooperate in a round, then she will be punished by the other players for the next $k$ rounds, where $k$ is finite and fixed.
\end{itemize}
It can be shown that for the repeated game for Prisoner's Dilemma, following Grim Trigger/Tit for Tat/Limited Punishment form Nash equilibrium. In the presence of such punishment strategies, whether cooperating actually becomes a Nash equilibrium depends on the discounting factor of the players. The discounting factor of the agents is an indicator of their patience. It determines whether agents will choose a short term profit by defecting, or choose to get consistent gains over a long period of time by cooperating in equilibrium. 

\begin{theorem}
A repeated game of the strategic form game $\Gamma$ will always have a Nash equilibrium
\end{theorem}

\begin{proof}
From \ref{thm:nash} we know that every strategic form game has a Nash equilibrium.

Clearly, the Nash equilibrium of $\Gamma$ will also be a Nash equilibrium of the repeated game of $\Gamma$. 
Thus, a repeated game of the strategic form game $\Gamma$ will always have a Nash equilibrium.
\end{proof}
Determining the subgame perfect equilibrium of a repeated game is a relatively simpler matter than doing so for general extensive games.
\begin{definition}[One Deviation Property \cite{osborne2004introduction}] No player can increase her payoff by changing her action at the start of any subgame in which she is the first mover, given the other players' strategies and the rest of her own strategy.
\end{definition}
\noindent This property is used to determine whether a strategy profile is a subgame perfect equilibrium for the given repeated game. 
\begin{theorem}[Subgame Perfect Equilibria in Repeated Games, \cite{osborne2004introduction}]
A strategy profile in a repeated game with a discount factor less than 1 is a subgame perfect equilibrium if and only if it satisfies the one deviation property. 
\end{theorem}

\noindent In our analysis, for a given pricing model and punishment model, we determine the subgame perfect equilibria. 

\subsection*{Punishment} When cooperation doesn't take place in a repeated game framework, a punishment aspect naturally comes up. In the model at hand, sellers cooperate by providing good quality products, and buyers do so by giving honest feedback. Our framework punishes buyers by means of the monitoring protocol and the smart contract deployed. Every infraction by a buyer, detected by the monitoring protocol (described in the next chapter), is penalized by a fee, which is the sum of the geometric progression with the first term as the maximum utility a buyer may receive in one round and the common ratio equal to $1-\nu$ with $\nu$ being a positive quantity close to $0$, agreed upon by all the agents in the market prior to its commencement. The fee is chosen in this manner to try to ensure that for providing dishonest feedback, the buyer will incur a loss greater than any utility she can receive from further purchasing on the market. We also ensure that our protocol does not detect dishonesty when the buyer has given honest feedback, thus removing any individual rationality issues.

Hence, in our framework, sellers do not actively punish the buyers, and we do not discuss any punishment strategies for sellers. The regulation protocols ensure that buyers will be honest. Buyers, on the other hand, can follow various punishment strategies, in order to penalize sellers for providing bad quality goods. To this effect, we study traditional punishment strategies, like tit for tat, which are implemented on an individual level by buyers for having received poor quality in some recent round. We also study punishment at a market-wide level as well, where a seller is punished by all buyers in the market, if his public rating drops below a threshold value. Additionally, we assume that all buyers follow the same punishment strategy. The next section analyzes this model under different pricing rules and punishment strategies.
\section{Analysis of Homogeneous Pricing} 
This is the simplest pricing model. Each seller sells her product at price $p$. In this model, buyers value a high-quality product at $v$ and a low-quality product at $0$. Thus, the payoff matrix of the buyer and seller from one single transaction becomes:

\begin{table}[h!]
\begin{center}
\begin{tabular}{|c|c|c|}
\hline
 & $H$ & $L$ \\
 \hline
Buy & $v-p,p-c$ & $-p,p$ \\
\hline
\end{tabular}
\caption{Payoff Matrix for Homogeneous Pricing \\With $v>p>c>0$}
\label{table:one}
\end{center}

\end{table}

\noindent This pricing rule and the rule for selecting a seller discussed in the previous chapter, ensure that:
\begin{itemize}
\item As long as a seller continues to provide high quality to all his buyers, he will not lose any buyers. The reason is that under homogeneous pricing, the price does not change, so the selection of a seller reduces to selecting a seller who will give the highest quality.
\item If all other sellers have only given poor quality, hence have $Q^t(s)=0$, a seller giving high-quality even once will be preferred henceforth, whenever he is not being punished.
\end{itemize}
We find that these hold for both the punishment models discussed below.

\subsection*{Punishment Model: Local Punishment}

In this model, we analyze traditionally studied punishment strategies of grim trigger, tit for tat and limited punishment. Under these strategies, a buyer will punish a seller who has given her low quality products by giving a bad feedback in that round and blacklisting for some rounds. The duration of the blacklisting is determined by the type of strategy chosen. Punishment is given to a seller $s$ in round $t$ by buyer b, only if $q_b^t (s)>0$. 

If all sellers have been blacklisted by a buyer in a round, the buyer will not make any purchase. Thus, in our model, a buyer will choose the seller who maximizes her expected utility, and is not being punished in that round. Thus, henceforth the strategy chosen by buyers to select the seller in a given round will not be explicitly considered, rather it will be considered as something that naturally follows from the punishment model, the quality received in previous interactions and the ratings. 

\subsection{Grim Trigger Strategies} Under this punishment strategy, on receiving poor quality, a buyer blacklists the seller for all future rounds. Thus, irrespective of the rating of other sellers, if a seller does not cooperate with a buyer once, the buyer will not return to the seller. For a seller to choose to cooperate with a seller, the utility of the seller for cooperating must be more than that for not cooperating. Let \textbf{$H^*$} denote the strategy that a seller gives out only  high quality products to all the buyers. Thus, his utility from \textbf{$H^*$} is the sum of the geometric progression with the first term as $p-c$ and common ratio $\sigma _s$, which is $\frac{p-c}{1-\sigma _s}$. The utility from not cooperating is $p$ alone, as the buyer does not return in future rounds. Thus, $s$ will follow \textbf{$H^*$} if \[ \frac{p-c}{1-\sigma _s}>p\] \[ \Rightarrow c<\sigma _s p\]

\noindent If $c>\sigma _sp$ then, by simply adding $p-c$ to both sides, $p>p-c +\sigma _s p$. Thus, a seller has no incentive to delay giving low-quality. This leads us to the following theorem: 

\begin{theorem}
Under grim trigger strategies, if $c<\sigma _s p$, then a seller will follow \textbf{$H^*$}, else will give only poor quality.
\end{theorem} 

\subsection{Tit for Tat Strategies} Under these strategies, blacklisting is for exactly one round. If all sellers are following \textbf{$H^*$}, then after the first round a seller will neither gain nor lose a buyer. After the first round, the value of $q_b^t (s)=1 \forall s,b$. To check if this is a subgame perfect equilibrium, we need only check if a single deviation for a seller will benefit them. Assuming all buyers are honest, if a seller $s$ gives a low-quality product to a buyer $b$, $b$ will punish $s$. As the feedback is honest, buyer $b$ will report $L$ and $s$ will lose all their customers in the next round. In the next round where $b$ does not blacklist $s$, as the value of $h_b (s)$ will have dipped, $b$ does not ever return to $s$. Thus, if at round $i$, $s$ has $n$ customers, then by giving only $b$ a low-quality product in this round, the seller will lose all the customers in the next round, nonetheless. Thus, seller $s$ will give out all the customers low-quality products. Hence, seller $s$ will deviate from \textbf{$H^*$} if:

\[ p>\frac{p-c}{1-\sigma _s}  \Rightarrow c<\sigma _s p\]

\noindent It is easy to see that high quality offerings are driven by (a) the $\frac{\sigma _s p}{c}$ ratio and (b) competition. Even if two sellers have $c<\sigma _s p$, the same reasoning will compel them to follow \textbf{$H^*$} in equilibrium. For a seller $s$, if $c>\sigma _sp$, there is no reason to delay giving poor quality, and hence $s$ will follow \textbf{$L^*$}. 
This leads us to the following theorem:

\begin{theorem}
If $c<\sigma _s p$, for at least two of the sellers, then the subgame perfect equilibrium for these sellers is to follow \textbf{$H^*$} and the others sellers to follow \textbf{$L^*$}.
\end{theorem} 

\noindent We now consider when would low-quality be a subgame perfect equilibrium. If all sellers follow \textbf{$L^*$}, then $Q^t (s)=0 \, \forall t>1$ and initially buyers cycle through all sellers, till $h_b ^t (s)=0\, \forall b,s$, after which each buyer randomly selects a seller in each round. Thus, a single deviation by seller $s$ to $H$ will result in all buyers buying from sellers $s$ in every round they are not punishing $s$. Thus, for all sellers playing \textbf{$L^*$} to be a subgame perfect equilibrium, we need that for all sellers: 

\[ p +\frac{n_B p\sigma _s}{n_S(1-\sigma _s)} > p-c + \frac{\sigma _s n_B p}{(1-\sigma _s^2)}\]

\[ \Rightarrow c>p\sigma _s n_B\left(\frac{1}{1-\sigma _s ^2}-\frac{1}{n_S(1-\sigma _s)}\right)\]

\noindent Thus, for \textbf{$L^*$} to be a subgame perfect equilibrium, the only condition is:

\[ \forall s, \,\,\, c>p\sigma _s n_B\left(\frac{1}{1-\sigma _s ^2}-\frac{1}{n_S(1-\sigma _S)}\right)\]

\noindent In this condition, \textbf{$L^*$} is not only an equilibrium strategy but also a dominant strategy. The sellers have no motivation to give even a single high-quality product, even if it means that they will get all the buyers henceforth. Hence, any seller, for whom $ c>p\sigma _sn_B\left(\frac{1}{1-\sigma _s ^2}-\frac{1}{n_S(1-\sigma _S)}\right)$, \textbf{$L^*$} is a dominant strategy. Therefore, we have the following theorem:

\begin{theorem}
If $c>p\sigma _sn_B\left(\frac{1}{1-\sigma _s ^2}-\frac{1}{n_S(1-\sigma _s)}\right)$, for a seller, then \textbf{$L^*$} is a dominant strategy.
\end{theorem} 

\noindent We now study the case when neither $H^*$ nor $L^*$ can be a subgame perfect equilibrium. 
\begin{theorem}
\label{thm:notfraud}
There is no pure strategy Nash equilibrium possible for homogeneous pricing other than $H^*$ and $L^*$.
\end{theorem}

\begin{proof}
Let us consider the setting where \[ \forall s \in S, \,\,\, p\sigma _s n_B\left(\frac{1}{1-\sigma _s ^2}-\frac{1}{n_S(1-\sigma _S)}\right) > c> \sigma _s p\]

\noindent If for all sellers, $\sigma _s$ is the same, then they will follow the same strategy in equilibrium. The analysis for this case generalizes to there being at least two sellers having the highest value of $\sigma_s$. It is easy to see that as these sellers are the most patient, all the buyers will be buying from these sellers, and the other sellers will be playing $L^*$. 

Consider, that there is no punishment. Thus, when all sellers are playing the same strategy, then buyers never switch from one seller to another. 

Having $\sigma _sp<c$ implies that the strategy followed cannot have only high-quality offerings, and  if $p\sigma _sn_B\left(\frac{1}{1-\sigma _s ^2}-\frac{1}{n_S(1-\sigma _S)}\right) > c$ implies that only low-quality will not be sustained. 

\subsubsection*{Necessary Conditions for Equilibrium}
Let \textbf{$(H^kL)^*$} be the periodic strategy where for $k$ rounds the seller gives high-quality then gives low-quality for one round and so on. To sustain \textbf{$(H^kl)^*$} as an equilibrium strategy there must be no incentive deviate from either $H$ or $L$ thus, to sustain this we need to ensure that there is no incentive to deviate from $H$.

The incentive to deviate from $H$ will be maximum after having given $L$ in the previous round. The utility at this point from a single buyer is: \[\frac{1}{1-\sigma _s ^{k+1}}\left(\frac{(p-c)(1-\sigma _s ^k)}{(1-\sigma_s)}+ \sigma _s ^{k+1}p\right)\]

\noindent On deviating, $s$ will lose all his buyers. Thus, if $s$ deviates, he will give all buyers low-quality. Consequently, his utility from a single buyer on deviating is $p$. Thus, to ensure that there is no incentive to deviate from $H$, we need

\begin{align*}
\frac{1}{1-\sigma _s ^{k+1}}\left(\frac{(p-c)(1-\sigma _s ^k)}{(1-\sigma_s)}+ \sigma _s ^{k+1}p\right)&>p \\
\Rightarrow  (p-c)(1-\sigma_s ^k) + \sigma _s ^{k+1}p(1-\sigma _s) &> p(1-\sigma _s ^{k+1})(1-\sigma_s)\\
\Rightarrow p(1 - \sigma_s^k +\sigma_s ^{k+1} - \sigma_s^{k+2} -((1-\sigma _s ^{k+1})(1-\sigma_s)) &>c(1-\sigma_s ^k)  
\end{align*}
On simplifying we get

\begin{equation}
c<\frac{p\sigma _s(1-\sigma ^{k-1} _s +2\sigma _s ^k(1-\sigma _s ))}{1-\sigma _s^k}
\end{equation}
We also need to ensure that there is no incentive to deviate from $L$. The utility on giving $L$ is \[p +\frac{\sigma_s}{1-\sigma_s^{k+1}}\left(\frac{(p-c)(1-\sigma _s ^k)}{(1-\sigma_s)}+ \sigma _s ^{k+1}p\right)\]
\noindent On deviating from $L$ for even one of his current customers, all buyers, who did not receive $L$ in this round will buy from $s$ in the next round. This incentive will be maximum when $s$ has only one buyer. Thus, we analyze this case. On deviating from $L$, the utility will be \[p-c +\frac{\sigma_sn_B}{1-\sigma_s^{k+1}}\left(\frac{(p-c)(1-\sigma _s ^k)}{(1-\sigma_s)}+ \sigma _s ^{k+1}p\right)\]

\noindent Thus, to ensure that there is no incentive to deviate from $L$ we need:

\begin{align*}
 c &> \frac{\sigma_s}{1-\sigma_s^{k+1}}\left(\frac{(p-c)(1-\sigma _s ^k)}{(1-\sigma_s)}+ \sigma _s ^{k+1}p\right) \left( n_B -1\right) \\
\Rightarrow c(1-\sigma_s^{k+1})(1-\sigma_s) &> \sigma_s ((p-c)(1-\sigma _s ^k)+ \sigma _s ^{k+1}p(1-\sigma_s))(n_B -1)
\end{align*}
On simplifying we get
\begin{equation}
c>\frac{p\sigma _s(1- \sigma _s ^k +\sigma _s ^{k+1}(1 -\sigma _s))(n_B -1)}{(1-\sigma_s^{k+1})(1-\sigma_s)+\sigma _s(1-\sigma _s ^{k})(n_B - 1)}
\end{equation}  
This can be rewritten as 
\begin{equation}
c>\frac{p\sigma _s(1- \sigma _s ^k +\sigma _s ^{k+1}(1 -\sigma _s)))}{x+\sigma _s(1-\sigma _s ^{k})}
\end{equation} 
where $x=\frac{(1-\sigma_s^{k+1})(1-\sigma_s)}{n_B -1}$

\subsubsection*{Infeasibility of Equilibrium Conditions}
Hence, to have $(H^kL)^*$ as an equilibrium strategy, we need
\begin{align}
\Rightarrow \frac{p\sigma _s(1-\sigma ^{k-1} _s +2\sigma _s ^k(1-\sigma _s ))}{1-\sigma _s^k}&>\frac{p\sigma _s(1- \sigma _s ^k +\sigma _s ^{k+1}(1 -\sigma _s)))}{x+\sigma _s(1-\sigma _s ^{k})} \\
\Rightarrow \frac{(1-\sigma ^{k-1} _s +2\sigma _s ^k(1-\sigma _s ))}{1-\sigma _s^k} &>\frac{(1- \sigma _s ^k +\sigma _s ^{k+1}(1 -\sigma _s))}{x+\sigma _s(1-\sigma _s ^{k})} \label{eq:imp}
\end{align}
Clearly,

\begin{align*}
(1-\sigma _s ^k)&= (1-\sigma _s)(1-\sigma _s ^k) + \sigma _s(1-\sigma _s ^k) \\
&>(1-\sigma _s)(1-\sigma _s ^{k+1}) + \sigma _s(1-\sigma _s ^k) \tag{$\because (1-\sigma _s ^{k+1})>(1-\sigma _s ^{k}) $} \\
&>\frac{(1-\sigma _s)(1-\sigma _s ^{k+1})}{n_B -1} +\sigma_s(1-\sigma _s ^k) \tag{$\because n_B\geq 3 \Rightarrow n_B-1\geq 2 >1 $}\\
&= x+ \sigma_s(1-\sigma _s ^k)
\end{align*}
Thus we have shown that the denominator in the LHS of \ref{eq:imp} will always be strictly greater than the denominator of the RHS.

\noindent Now we consider the difference between the numerator of the RHS and that of the LHS of \ref{eq:imp}
\begin{align*}
&(1- \sigma _s ^k +\sigma _s ^{k+1}(1 -\sigma _s)) - (1-\sigma ^{k-1} _s +2\sigma _s ^k(1-\sigma _s )) \\
=& - \sigma _s ^k +\sigma _s ^{k+1}(1 -\sigma _s)) +\sigma ^{k-1} _s -2\sigma _s ^k(1-\sigma _s ) \\
=& \sigma_s ^{k-1}(1+ \sigma _s^2(1 -\sigma _s)) - \sigma _s ^k(3-2\sigma_s) \\
=& \sigma_s ^{k-1}( 1+ \sigma _s^2(1 -\sigma _s) - \sigma _s(3-2\sigma_s) \\
=& \sigma_s ^{k-1} (1+ 3\sigma^2-3\sigma-\sigma^3) \\
=& \sigma_s ^{k-1}(1-\sigma_s)^3 
\end{align*}
Now, as $\sigma_s \in (0,1)$, this value will always be greater than 0.
Thus, we have that the numerator in the LHS of \ref{eq:imp} will always be strictly less than the numerator of the RHS. As a result, \ref{eq:imp} will never be satisfied.

To reduce the incentive to deviate from $H$, we need a higher $\frac{\sigma _s p}{c}$ ratio, and to reduce the incentive to deviate from $L$ we need a lower $\frac{\sigma _s p}{c}$ ratio. Consequently, $(H^kL)^*$ will never be a subgame perfect equilibrium, for any value of $k\in \mathbb{N}\backslash{0}$. 

\subsubsection*{Considering Other Strategies}
Here we have considered a setting, without punishment. Having punishment of any form will only increase the incentive to deviate from $L$, and decrease the incentive to deviate from $H$. Thus, even with punishment, $(H^kL)^*$ will never be a subgame perfect equilibrium, for any value of $k\in \mathbb{N}\backslash{0}$.   
Increasing the number of low-quality given, will simply increase the incentive to deviate from $L$.

Let us now consider a setting where the highest value of $\sigma _s$ is achieved by exactly one seller, say $s_1$. 
Let $s_1$ play strategy $(H^kL)^*$, with $k\in \mathbb{N}\backslash \{0\}$, such that $k$ is more than what the other sellers can sustain.

Thus, the best response of all the other sellers is to play $L^*$. But the best response of $s_1$ to all other sellers playing $L^*$ is to reduce the amount of high-quality. In this case the best response of the seller with the second highest value of $\sigma_s$, call this seller $s_2$, would be to start giving $H$ so as to have  $s_1$'s customers switch to $s_2$, forcing $s_1$ to increase the amount of high-quality back to $k$. 

Thus, there is no pure strategy equilibrium in this case.  

Hence, there is no pure strategy subgame perfect equilibrium when:

\[ \forall s \in S , \,\,\, p\sigma _s n_B\left(\frac{1}{1-\sigma _s ^2}-\frac{1}{n_S(1-\sigma _S)}\right) > c> \sigma _s p\] 

\end{proof}

\noindent Along with the characterization of pure strategy subgame perfect equilibria, we also infer that \textbf{competition} and \textbf{profit margin} are major driving factors in the quality offered by sellers. If at least $n_S -1$ sellers have $c>\sigma _sp$ and exactly one seller has the highest $\frac{\sigma _sp}{c}$ ratio, then this seller has a monopoly, and will follow the strategy that maximizes his utility. This strategy also ensures that all buyers in the market return to him, whenever they are not punishing him. This case is neither particularly interesting nor practical. We now study the effect of increasing the duration of blacklisting.


\subsection{Limited Punishment Strategies}

In limited punishment, punishment is given by buyer $b$ for receiving low-quality from a seller $s$ for $\alpha$ rounds. This implies, that the buyer will give low feedback in that very round and for the next $\alpha$ rounds, the buyer will blacklist seller $s$. 

\begin{theorem}
If $c<\sigma _s p$, for at least two of the sellers, then the subgame perfect equilibrium for these sellers is to follow \textbf{$H^*$} and the others sellers to follow \textbf{$L^*$}.
\end{theorem} 

\begin{proof}
Similar to tit for tat, the utility from playing $H^*$ in any given round is $\frac{p-c}{1-\sigma_s}$.
On deviating, the seller will lose all his customers. Hence, the utility on deviating is $p$.
Thus, for $H^*$ to be a subgame perfect equilibrium strategy for $s$, we need \[c<\sigma_sp\] When at least two sellers have $c<\sigma_sp$, then all other sellers will play $L^*$. 
\end{proof}
\noindent We now characterize when giving low quality will be a SGPE.
\begin{theorem}
If $c>p\sigma _sn_B\left(\frac{1}{1-\sigma _s ^{\alpha+1}}-\frac{1}{n_S(1-\sigma _s)}\right)$, for all sellers, then all sellers playing  \textbf{$L^*$}is a subgame perfect equilibrium.
\end{theorem} 
\begin{proof}
All sellers following \textbf{$L^*$} and all buyers giving honest feedback will be a subgame perfect equilibrium if:
\[ p +\frac{n_B p\sigma _s}{n_S(1-\sigma _s)} > p-c + \frac{\sigma _sn_Bp}{(1-\sigma _s^{\alpha+1})}\] 
\[ \Rightarrow c>p\sigma _s n_B\left(\frac{1}{1-\sigma _s ^{\alpha+1}}-\frac{1}{n_S(1-\sigma _s)}\right)\]
\end{proof}

\begin{theorem}
If $c>p\sigma _sn_B\left(\frac{1}{1-\sigma _s ^{\alpha+1}}-\frac{1}{n_S(1-\sigma _s)}\right)$, for a seller, then \textbf{$L^*$} is a dominant strategy.
\end{theorem} 

\begin{proof}
The sellers have no motivation to give even a single high-quality product, even if it means that they will get all the buyers henceforth. Hence, any seller, for whom $ c>p\sigma _sn_B\left(\frac{1}{1-\sigma _s ^{\alpha+1}}-\frac{1}{n_S(1-\sigma _s)}\right)$, \textbf{$L^*$} is a dominant strategy.
\end{proof}

\noindent From Theorem \ref{thm:notfraud}, we have that there can be no other pure strategy SGPE. 

Clearly, increasing the duration of the blacklisting reduces the incentive to provide low-quality, even in the case of monopolies, where at least $n_S -1$ sellers have $c>\sigma _s p$. This analysis completes the characterization of subgame perfect equilibria under typical punishment strategies for repeated games. We now explore punishment strategies implemented market-wide.

\subsection{Punishment Model: Threshold Punishment}
In this punishment model, 
punishment is no longer given for every infraction, by a single buyer. Rather, 
the market as a whole blacklists a seller if their public perception falls below a threshold value for $\alpha$ rounds. The reason for exploring this model is, that in the pure strategy equilibria in the local punishment model, a single infraction , despite having given consistent high-quality earlier, necessitated that the buyer switch to another seller, even though the other seller is behaving identically. This is not very practical.

The selection of the seller is still on the basis of maximization of expected utility, based on their own estimated likelihood of getting a high-quality product. A seller is blacklisted only if the value of $Q^t(s)$ is less than the specified threshold. A blacklisted seller is reintroduced to the market after $\alpha$ rounds with the threshold value as his rating. 

For homogeneous pricing, we set the threshold value as that below which the expected utility would be negative. Thus, the threshold for homogeneous pricing is $p/v$. Analysis similar to that of the always punish model helps us find the subgame perfect equilibria.

\begin{theorem}
If $c<\sigma _s p$, for at least two of the sellers, then the subgame perfect equilibrium for these sellers is to follow \textbf{$H^*$} and the others sellers to follow \textbf{$L^*$}.
\end{theorem} 
\begin{proof}
Similar to local punishment, the utility from playing $H^*$ in any given round is $\frac{p-c}{1-\sigma_s}$.On deviating, the seller will lose all his customers. Hence, the utility on deviating is $p$.
Thus, for $H^*$ to be a subgame perfect equilibrium strategy for $s$, we need \[c<\sigma_sp\] When at least two sellers have $c<\sigma_sp$, then all other sellers will play $L^*$. 

\end{proof}

\begin{theorem}
If $c>\frac{\sigma _s pn_B}{(1-\sigma_s^{\alpha+1})}\left( 1-\frac{\sigma _s ^{\alpha +1}}{n_S}\right)$, for a seller, then \textbf{$L^*$} is a dominant strategy.
\end{theorem} 

\begin{proof}
All sellers following \textbf{$L^*$} and all buyers giving honest feedback will be a subgame perfect equilibrium if:

\begin{align*}
p +\frac{n_B p\sigma _s ^{\alpha +2}}{n_S(1-\sigma _s ^{\alpha +1})} &> p-c + \frac{\sigma _sn_Bp}{(1-\sigma _s^{\alpha+1})}\\ 
 \Rightarrow c&>\frac{\sigma _s pn_B}{(1-\sigma_s^{\alpha+1})}\left(1-\frac{\sigma_s^{\alpha+2}}{n_S(1-\sigma _s)}\right)
\end{align*}

\end{proof}

\noindent From Theorem \ref{thm:notfraud} we have that there is no other SGPE. 

Let us now compare the bounds established by limited punishment and threshold punishment to find which model disincentivizes low quality better. 

\begin{theorem}
Threshold Punishment model disincenitvizes low-quality offerings better than Local Punishment.
\end{theorem}

\begin{proof}
For threshold punishment we need $c>\frac{\sigma _s pn_B}{(1-\sigma_s^{\alpha+1})}\left( 1-\frac{\sigma _s ^{\alpha +1}}{n_S}\right)$, for  \textbf{$L^*$} to be a dominant strategy. For limited punishment, we need  $c>p\sigma _sn_B\left(\frac{1}{1-\sigma _s ^{\alpha+1}}-\frac{1}{n_S(1-\sigma _s)}\right)$.

Consider,

\begin{align*}
&\frac{\sigma _s pn_B}{(1-\sigma_s^{\alpha+1})}\left( 1-\frac{\sigma _s ^{\alpha +1}}{n_S}\right)-p\sigma _sn_B\left(\frac{1}{1-\sigma _s ^{\alpha+1}}-\frac{1}{n_S(1-\sigma _s)}\right) \\
=& \sigma _s p n_B \left(\frac{1}{(1-\sigma_s^{\alpha+1})}\left( 1-\frac{\sigma _s ^{\alpha +1}}{n_S}\right)-\left(\frac{1}{1-\sigma _s ^{\alpha+1}}-\frac{1}{n_S(1-\sigma _s)}\right) \right)\\
=& \frac{\sigma _s p n_B}{(1-\sigma_s^{\alpha+1})(n_S(1-\sigma_s))} \left(n_S(1-\sigma_s) -\sigma_s^{\alpha+2} -(1-\sigma^{\alpha+1})\left(\frac{n_S(1-\sigma_s)}{1-\sigma _s ^{\alpha+1}}-1  \right)\right)\\
=& \frac{\sigma _s p n_B}{(1-\sigma_s^{\alpha+1})(1-\sigma_s^{\alpha +1})(n_S(1-\sigma_s))} \left(n_S(1-\sigma_s) -\sigma_s^{\alpha+2} -(1-\sigma^{\alpha+1})(n_S(1-\sigma_s) -1+\sigma_s^{\alpha+1}) \right) \\
=&\frac{\sigma _s p n_B}{(1-\sigma_s^{\alpha+1})(1-\sigma_s^{\alpha +1})(n_S(1-\sigma_s))} \left(1- \sigma_s^{\alpha+1}+\sigma_s^{\alpha+2}-\sigma_s^{2\alpha+2}\right) \\
>&0 \tag{$\because \frac{\sigma _s p n_B}{(1-\sigma_s^{\alpha+1})(1-\sigma_s^{\alpha +1})(n_S(1-\sigma_s))}>0$ and $1- \sigma_s^{\alpha+1}+\sigma_s^{\alpha+2}-\sigma_s^{2\alpha+2}>0$}
\end{align*}
Thus, we infer that threshold punishment model is comparatively better for disincentivizing poor quality. Consequently, we conclude that for homogeneous pricing, threshold punishment is the most conducive to inducing high quality offerings from sellers.
\end{proof}

\noindent It is also important to note that while all sellers selling at the same price is a useful building block for further analysis, it is not very pragmatic. We do not see different qualities of the same product being traded at the same price. Thus, we now explore heterogeneous pricing. 

\section{Analysis with Binary Pricing }
We now take into consideration real-world markets, where sellers of the same product may either belong to a high-price category or low-price category. The price category to which the seller belongs to is often an indicator of the quality of the product provided by him. We abstract this situation into the binary pricing model as follows. 

We assume that each buyer has value $v_L$ and $v_H$ for low and high-quality respectively, with $v_H>v_L >0$. 

In this model, goods can be sold at either $p_H$ or $p_L$, with the cost of producing a high and a low-quality good as $c$ and $0$ respectively. 

We assume
\begin{itemize}
\item $ p_H>p_H-c>p_L>p_L-c>0$ 
\item $v_H-p_L>v_H-p_H>v_L-p_L>0>v_L-p_H$.
\end{itemize}

\noindent Let $S_H$ and $S_L$ be the set of sellers selling at $p_H$ and $p_L$ respectively. Thus, $S=S_H \cup S_L$.
Let $n_H=|S_H|$ and $n_L=|S_L|$.

We consider only the threshold punishment model as it tends to induce the \textbf{$L^*$} equilibrium less than the always punish model, as shown in the previous section. 

The threshold for sellers in $S_H$ is $p_H/v_H$, similar to the homogeneous pricing model, the threshold for sellers in $S_L$ is $\epsilon$ such that $0<\epsilon<<1$ and is introduced to demotivate \textbf{$L^*$} becoming an equilibrium strategy.

Within binary pricing we consider two pricing models:
\begin{enumerate}
\item Non-adaptive binary pricing: The sets $S_H$ and $S_L$ do not change and
\item Adaptive binary pricing: The prices charged by the sellers depend on their public perception.
\end{enumerate}
We show that given the parameters of the system, both of these pricing rules reduce to different cases of homogeneous pricing.

\subsection{Non-adaptive Binary Pricing}
This models markets where the prices offered by a seller are consistent over time. On analysis, we find that this pricing rule reduces to different instances of the homogeneous pricing model, under different conditions. 

Throughout the game, sellers in $S_H$ charge $p_H$, and those in $S_L$ charge $p_L$. In this model, we assume initial perception of a seller, $\xi$ to be the same irrespective of the price charged. Thus, initially, as the estimated likelihood of getting a high-quality product is the same for all sellers, in the first round, a buyer randomly chooses a seller in $S_L$.

Buyers expect a utility of $\xi v_H + (1-\xi )v_L -p_H$ from sellers in $S_H$. Thus, as long as at least one seller in $S_L$ has expected utility at least $\xi v_H + (1-\xi )v_L -p_H$ then buyers never buy from a seller in $S_H$.

If $c<\sigma _sp_L$ for at least two sellers in $S_L$, then the buyers get a utility of $v_H-p_L$ from these sellers and will continue to buy from these sellers. Similarly, any equilibrium which gives a utility of at least $\xi v_H + (1-\xi )v_L -p_H$ can be sustained, and no buyer will ever buy from a seller in $S_H$. 

If a seller in $S_L$ cannot sustain a strategy that gives the buyers a utility of at least $\xi v_H + (1-\xi )v_L - p_H$ then they give out $L$ from the very first round. In the case that all sellers in $S_L$ are in this state, only then buyers will proceed to buy from sellers in $S_H$. 

Thus, once buyers start buying from sellers in $S_H$ as long as at least one of the sellers can give an expected utility of at least $v_L-p_L$, buyers will continue to buy from sellers in $S_H$ else will switch back to sellers in $S_L$ and never again buy from sellers in $S_H$. All sellers who cannot sustain a strategy which guarantees an expected utility of $v_L - p_L$ and avoids going into isolation will follow \textbf{$L^*$}.

If the buyers switch from buying from sellers in $S_H$ to buying from sellers in $S_L$, the game reduces to that of the homogeneous pricing model with threshold punishment.

The results are summarized in the following table

\begin{table}[h!]
\centering
\label{table:nonadbp}
\begin{tabular}{|c|c|}
\hline
Condition & Homogeneous Pricing Parameters \\
\hline
$\exists s \in S_L$ can sustain $\geq \xi v_H + (1-\xi )v_L - p_H$ & $S=S_L$, $p=p_L$ \\
\hline 
$\exists s \in S_H$ can sustain $\geq v_L-p_L$ & $S=S_H$, $p=p_H$ \\
\hline
$\nexists s \in S_H$ can sustain $\geq v_L-p_L$ & $S=S_L$, $p=p_L$ \\
\hline
\end{tabular}
\caption{Results of Non-Adaptive Binary Pricing}
\end{table}
\subsection{Adaptive Binary Pricing}
We now model markets where prices are not consistent over time, such as with new technology. Initially, all sellers charge a high price in order to indicate quality, but may later reduce the price. We abstract this situation into an adaptive binary pricing model. The sellers switch their states based on their rating. All sellers start at price $p_H$. If $Q^t(s)<\frac{p_H}{v_H}$ the seller is isolated for $\alpha$ rounds, after which the seller joins $S_L$ and charges $p_L$, with $Q^t(s)=\max (\epsilon,Q^{t-\alpha-1}(s))$. The seller is upgraded to price $p_H$ once the estimated utility, based on the rating crosses $v_H-p_H$, thus when $Q^t(s)\geq 1-\frac{p_H-p_L}{v_H-v_L}$. The reason for having all sellers start at the same price, is that if sellers cannot sustain giving high-quality at a higher price, they will not be able to do so at a lower price.
\begin{figure}[h!]
\label{fig:binp}
\hspace{0.15\linewidth}
\includegraphics[width=0.6\linewidth]{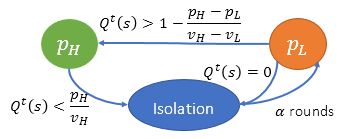}
\caption{Adaptive Binary Pricing: Price update Rule}
\end{figure}

\noindent All those sellers who cannot sustain a strategy which avoids isolation will follow \textbf{$L^*$}. If there is at least one seller who can avoid isolation, then the game reduces to homogeneous pricing at $p_H$ with the sellers who can avoid isolation. 

If there are no such sellers, then the game reduces to that of homogeneous pricing at price $p_L$ with all the sellers.  Thus we have,

\begin{table}[h!]
\centering
\label{table:adbp}
\begin{tabular}{|c|c|}
\hline
Condition & Homogeneous Pricing Parameters \\
\hline
$\exists s \in S$ who can avoid isolation & $S=S_H$, $p=p_H$ \\
\hline 
$\nexists s \in S$ who can avoid isolation & $S=S_L$, $p=p_L$ \\
\hline
\end{tabular}
\caption{Results of Adaptive Binary Pricing}
\end{table}
\noindent As both the pricing rules in binary pricing reduce to various cases of homogeneous pricing, we choose not to explore other discrete pricing models with more prices as such models will not achieve anything more than exploding the state space. 

\noindent A consequence of the two pricing rules reducing to different cases of homogeneous pricing is:
\begin{itemize}
\item As long as a seller continues to provide high quality to all his buyers, he will not lose any buyers.
\item If all other sellers are only giving poor quality, a seller giving high-quality even once will be preferred henceforth, whenever he is not being punished.
\end{itemize}

\noindent This would not be found in a continuous pricing model, where with even small changes in rating, the price charged changes. 

\section{Analysis with Continuous Pricing}
We assume that each buyer has value $v_L$ and $v_H$ for low and high-quality respectively, with $v_H>v_L >0$.

In this model the price charged by a seller is contingent on the public perception of the seller. We choose to linearly scale the public perception of a seller to the range $[p_L,p_H]$ to determine the price. That is, seller $s$ will charge in round $t$: $$p_L +(p_H-p_L)Q^t(s)$$
Under this model, we observe immediate changes in prices even with minor changes in the public perception, bringing about behavior not found in the pricing models discussed earlier. The behavior shown by a buyer, $b$ in this model depends also on her type, $\theta _b$, which was not observed in the preceding analysis. Contrary to previous results, in the continuous pricing model, we find that preferring a seller with rating 1 over one with lesser rating is not necessary. Nor is it the case that, if all other sellers were only giving low quality, to all their buyers, and a seller were to give high quality to even a fraction of his buyers, in the next round, all buyers would choose to buy from him. 

\begin{theorem}
If a buyer $b$ has to choose between sellers $s_1$ and $s_2$, in round $t$, where $h_b^t(s_1)=Q^t(s_1)=1$ and $h_b^t(s_2)=1-\phi$ and $Q^t(s_2)=1-\epsilon$, with $\phi\geq 0$ and $\epsilon > 0$, she will choose $s_1$ if $$\theta _b\left(1-\frac{\phi}{\epsilon}\right) < 1 -\left(\frac{p_H-p_L}{v_H-v_L}\right)$$  
\end{theorem}

\begin{proof}
It is easy to see that $q_b ^t(s_1)= 1$ and $q_b ^t(s_2)= \theta _b(1-\phi )+ ((1-\theta _b)(1-\epsilon ))=1-(\epsilon (1- \theta _b) +\theta _b\phi)$.
Let $\beta = \epsilon (1- \theta _b) +\theta _b\phi$

The price charged by $s_1$ is $p_H$ and by $s_2$ is $p_L +(1-\epsilon)(p_H-p_L)=p_H - \epsilon(p_H -p_L)$
If $b$ must choose $s_1$ over $s_2$, the utility $b$ expects from $s_1$ must be greater than what she expects from $s_2$. Thus, we need:
\[ v_H - p_H > (1-\beta )v_H +\beta v_L -p_H+ \epsilon(p_H-p_L)\] 
\[ \implies \beta (v_H - v_L) -\epsilon(p_H-p_L)>0\] 
\[ \implies (\epsilon (1- \theta _b) + \theta _b \phi )(v_H-v_L) -\epsilon (p_H-p_L)>0 \] 
\[ \implies \frac{\epsilon(p_H-p_L)}{(v_H-v_L)}<\epsilon (1-\theta _b) +\theta _b \phi\]
\[ \implies \frac{(p_H-p_L)}{(v_H-v_L)}< (1-\theta _b) +\theta _b \frac{\phi}{\epsilon}\]
\[ \therefore \theta _b\left(1-\frac{\phi}{\epsilon}\right) < 1 -\left(\frac{p_H-p_L}{v_H-v_L}\right)\] 
\end{proof}

\begin{theorem}
If a buyer $b$ has to choose between sellers $s_1$ and $s_2$, in round $t$, where $h_b^t(s_1)=Q^t(s_1)=0$ and $h_b^t(s_2)=\phi$ and $Q^t(s_2)=\epsilon$, with $phi\geq 0$ and $\epsilon > 0$, she will choose $s_2$ if $$\theta _b\left(1-\frac{\phi}{\epsilon}\right) < 1 -\left(\frac{p_H-p_L}{v_H-v_L}\right)$$  
\end{theorem}

\begin{proof}
It is easy to see  that $q_b ^t(s_1)= 0$ and $q_b ^t(s_2)= \theta _b\phi + (1-\theta _b)\epsilon$.
Let $\beta = \epsilon (1- \theta _b) +\theta _b\phi$. The price charged by $s_1$ is $p_L$ and by $s_2$ is $p_L +(\epsilon)(p_H-p_L)$. If $b$ must choose $s_2$ over $s_1$, the utility $b$ expects from $s_2$ must be greater than what she expects from $s_1$. Thus, we need, 
$ v_L - p_L < (1-\beta )v_H +\beta v_L -p_H+ \epsilon(p_H-p_L)$.  
From analysis similar to above, we get:
\[ \theta _b\left(1-\frac{\phi}{\epsilon}\right) < 1 -\left(\frac{p_H-p_L}{v_H-v_L}\right)\]
\end{proof}
\noindent These results capture following 
\begin{itemize}
\item If the buyer does not give much importance to public perception, a change in the public perception is more significant in the change in price than in quality.
\item If the value of $\frac{p_H-p_L}{v_H-v_L}$ is very close to $1$, then a buyer is almost indifferent to quality and would prefer to save on the price.
\item If the change in the quality delivered by the seller to the buyer as at least as much as that in the public perception, then the buyer will always prefer better quality
\end{itemize}

\noindent Let us now characterize how buyers will choose between two sellers in this pricing model.

Given two sellers $s_1$ and $s_2$ and a buyer $b$.

Let $h^t_b(s_1)=\phi _1$ and $h^t_b(s_2)=\phi _2$, and $Q^t(s_1)=\epsilon_1$ and $Q^t(s_2)=\epsilon_2$.

Thus, $q^t_b(s_1)=\theta_b \phi_1 +(1-\theta_b)\epsilon_1$ and $q^t_b(s_2)=\theta_b \phi_2 +(1-\theta_b)\epsilon_2$.

Let $E_b^t(s)$ denote the utility $b$ expects from seller $s$ in round $t$.

\begin{align}
\therefore E_b^t(s_1) &= (\theta_b\phi_1+(1-\theta_b)\epsilon_1)v_H +(1-\theta_b\phi_1 -(1-\theta_B)\epsilon_1)v_L -\epsilon_1p_H-(1-\epsilon_1)p_L \\
&= (v_H -v_L)(\theta_b\phi_1+(1-\theta_b)\epsilon_1)+(v_L-p_L)-\epsilon_1(p_H-p_L) \label{eq:exp}
\end{align}

\noindent Similarly 

\begin{equation}
E_b^t(s_2)= (v_H -v_L)(\theta_b\phi_2+(1-\theta_b)\epsilon_2)+(v_L-p_L)-\epsilon_2(p_H-p_L) 
\end{equation}

\noindent For $b$ to choose $s_1$ over $s_2$, we need

\[E_b^t(s_1)>E_b^t(s_2) \]
\[(v_H -v_L)(\theta_b(\phi_1-\phi_2)+(1-\theta_b)(\epsilon_1-\epsilon_2))-(\epsilon_1-\epsilon_2)(p_H-p_L) >0 \tag{On simplifying}\]

\noindent Let $\epsilon_1>\epsilon_2$, then on dividing by $(\epsilon_1 -\epsilon_2)$, we have

\[(v_H-v_L)\left(\theta_b\left(\frac{\phi_1-\phi_2}{\epsilon_1-\epsilon_2}-1\right)+1\right)-(p_H-p_L)>0\]
\[\Rightarrow 1- \frac{p_H-p_L}{v_H-v_L}>\theta_b \left(1-\left(\frac{\phi_1-\phi_2}{\epsilon_1-\epsilon_2}\right)\right)\]

\noindent Thus, we have that if $\epsilon_1>\epsilon_2$, $b$ will choose $s_1$ over $s_2$ if $ 1- \frac{p_H-p_L}{v_H-v_L}>\theta_b \left(1-\left(\frac{\phi_1-\phi_2}{\epsilon_1-\epsilon_2}\right)\right)$

Else, if  $\epsilon_1<\epsilon_2$, $b$ will choose $s_1$ over $s_2$ if $ 1- \frac{p_H-p_L}{v_H-v_L}<\theta_b \left(1-\left(\frac{\phi_1-\phi_2}{\epsilon_1-\epsilon_2}\right)\right)$

Clearly, if $\epsilon_1=\epsilon_2$ $b$ will choose $s_1$ over $s_2$ if $\phi_1>\phi_2$.\\

\noindent Let $\beta= \frac{p_H-p_L}{v_H-v_L}$. $\beta n_B$ is the expected number of customers for whom $\theta _b > 1 -\left(\frac{p_H-p_L}{v_H-v_L}\right)$, and thus may value price over the quality.

Based on this analysis, we derive the following table of the expected number of buyers in the following round based on the strategy chosen in the current round and previous history, given there is exactly one other seller playing $H^*$. Let $b^t(s)$ be the number of buyers seller $s$ has in round $t$ 

\begin{table}[ht]
\centering
\begin{tabular}{|c|c|c|}
\hline
History&Strategy&Expected Customers in $t+1$\\
\hline
Anything &$H^*$ &$b^t(s)$\\
$H^*$ & $H^kL^{b^t(s)-k}$& $\beta(n_b+k-b^t(s))$ \\
Not $H^*$ &$H^kL^{b^t(s)-k}$&$k$\\
\hline
\end{tabular}
\end{table}
\noindent We observe that even for buyers, for whom $\theta _b > 1 -\left(\frac{p_H-p_L}{v_H-v_L}\right)$, having received low-quality from a seller, the buyer never return to this seller, as long as there is a seller giving better quality. The behavior described above also severely limits the analysis possible in this pricing model. No closed form expressions are found for the analysis for equilibrium behavior. This concludes our exploration of different pricing models. We now list the details of our regulation protocols which form the backbone of our model.

\subsection*{Implementation}
The analysis presented in this chapter relies on the existence of a B2B market platform as described in the previous chapter. The assumptions made are supported by available blockchain technology. A typical blockchain fabric like Hyperledger Fabric (\cite{cachin2016architecture}) can be used to implement the proposed system. It has a concept of channels where data and ledger is shared across participants in the channel. 

In the case studied in this thesis, a particular seller and buyer pair will be part of one private channel, that is a B2B network. However, a buyer or a seller may be part of multiple such channels, with different agents. Transactional privacy is maintained across blockchain channels, that is, across different B2B networks. All data related to the channel, such as channel information, member, transaction, et cetera, is not accessible by anyone not explicitly granted access to that channel. 
Further, a ``meta'' channel, consisting of all buyers and sellers, will help facilitate computation of reliable ratings.
On top of these multiple channels, the deployment of inter-channel smart contracts is possible. Such smart contracts, which enforce pricing rules and punishment strategies, and compute reliable ratings, together form the B2B market platform.

\section{Conclusions}
Repeated games enable us to model punishment. In our case, there is a monitoring protocol in place to ensure that the buyer is honest, thus punishment only applies to sellers. The local punishment model captures traditional punishment strategies, where punishment is given for every single infraction. The threshold punishment model gives punishment is given market-wide, on the rating having dropped below a certain threshold. We find that over local punishment, threshold punishment is able to better disincentivize poor-quality offerings from the sellers. An advantage of the threshold punishment rule is that it can be enforced using smart contracts. In fact, as discussed earlier, the very purpose of the analysis presented in this chapter is to be able to design smart contracts that enforce conditions which best incentivize the enterprise sellers to deliver high-quality. The smart contracts deployed can enforce particular seller selection rules, punishment strategies as well as regulate the prices charged by a seller. Consequently, this chapter presents the various equilibria that surface under different pricing rules.

While the case where all sellers sell at the same price may seem unrealistic, it is found that whenever the price charged by a seller does not change with a small change in the rating, the system reduces to a case of homogeneous pricing. When, the price charged by a seller does in fact change with even minute changes in the seller's rating, buyers may switch to a seller with a lower rating than the one they were previously buying from. This may be a consequence of the buyer having little value for the public perception of a seller. However, if any buyer were to receive low-quality herself, she will never return to that seller, as long as there is another seller having a rating at least as high. The protocol deployed to actually compute these ratings from the buyers' feedback vectors is detailed in the following chapter.

\chapter{Cryptographic Regulation Protocols}
\begin{quote}
{\em This chapter elaborates on the cryptographic regulation protocols deployed on the platform that fulfill the privacy requirements of B2B platform. We describe in particular, two subprotocols, namely the Monitoring Protocol and the Public Perception Protocol which facilitate the interactions discussed previously using public key schemes and homomorphic encryption.}
\end{quote}

\section{Cryptographic Protocols: Some Preliminaries}
Since the beginning of civilization, but even more so since the advent of the Internet, all entities are associated with information. Some information can be made publicly available whereas some must be deemed private. With online computations gaining prominence, often the need arises to communicate securely and compute on sensitive and private data of various entities. Clearly non-trivial, the goal of secure communication is well stated and understood.
Encryption schemes play a vital role in securing data transfer. Efficient encryption schemes have been designed over the years to tackle the problem of secure communication. However, cryptographic community these days focuses on problems that are well beyond the problem of securing the data in transit. Cryptography has evolved over the years to tackle the problem of keeping the data private as well as performing computation on it. 

In essence, the field of cryptography can be broadly classified into two classes: symmetric key cryptography and public key cryptography. Particularly in the encryption paradigm, symmetric key algorithms use a single key for both encryption of a message and its decryption whereas public key algorithms allow use of two keys: a public-key which is distributed widely is used to encrypt a message and a corresponding secret key, known only to the owner is used to decrypt the ciphertext to obtain the underlying message. For communication involving large number of parties, a public-key based approach seems more plausible since symmetric key based schemes require exchange of pairwise keys between every pair of parties.

In this section, we describe the necessary primitives required for the construction of our cryptographic protocol that aims at secure computation of public perception vector and monitoring of feedback. We assume pairwise authentic channels are available between the parties for communication.

\noindent \textbf{Public Key Encryption}: A public key encryption scheme is defined using the tuple of probabilistic polynomial-time algorithms key generation, encryption and decryption indicated as (\textbf{Gen}, \textbf{Enc}, \textbf{Dec}) respectively satisfying:
\begin{itemize}
\item \textbf{Gen}: ($pk$, $sk$) $\leftarrow$ \textbf{Gen}$(1^\kappa)$. The key generation algorithm Gen takes as input security parameter $\kappa$ and generates a key pair ($pk$, $sk$) where $pk$ is the public key and $sk$ is the secret key. This is a randomized algorithm. 
\item \textbf{Enc}: $c$ $\leftarrow$ \textbf{Enc}$(m,pk)$. The encryption algorithm Enc takes as input the plaintext message $m \in \mathcal{M}$, where $\mathcal{M}$ is the message space and public key, $pk$ to generate a ciphertext $c \in \mathcal{C}$, where $\mathcal{C}$ is the ciphertext space.This is also a randomized algorithm.
\item  \textbf{Dec}: $m$ = \textbf{Dec}$(c,sk)$. The decryption algortihm Dec takes as input the ciphertext $c$ and secret key $sk$ to obtain the corresponding plaintext $m$ or special symbol $\perp$ denoting decryption failure. This is usually a deterministic algorithm.
\end{itemize}

\noindent \textbf{Homomorphic Encryption}: An homomorphic encryption scheme \cite{Gentry09} $\pi$ is defined using a tuple of four algorithms (\textbf{KeyGen}, \textbf{Encrypt}, \textbf{Decrypt}, \textbf{Evaluate}) with the first three algorithms performing usual operations and the algorithm \textbf{Evaluate} takes as input the public key $pk$, a circuit $C$ from a set of permitted circuits and a tuple of ciphertexts $c$=$(c_1, c_2, ..., c_n)$ corresponding to plaintexts $(p_1, p_2, ..., p_n)$ to obtain a ciphertext $c'$ which on decryption produces the plaintext $p'$ = $C$$(p_1, p_2, ..., p_n)$. We denote it as:
\[c' \leftarrow \textbf{Evaluate}(pk, C, c) \Rightarrow C(p_1, p_2, ..., p_n) = \textbf{Decrypt}(sk, c')\]

\noindent \textbf{Fully Homomorphic Encryption (FHE)}: The scheme $\pi$ is fully-\\homomorphic if it is homomorphic for all circuits \cite{Gentry09}.
Given a set of ciphertexts $(c_1, c_2, ..., c_n)$ encrypted under a public key, an encryption that supports any joint computation on these ciphertexts to obtain a ciphertext $C = f(c_1, c_2, ..., c_n)$, where $f$ is a polytime computable function is called a fully homomorphic encryption. No information about the individual ciphertexts, $C$ or any intermediate values should be leaked during computation. By default, homomorphic cryptosystems are malleable.\\

\noindent \textbf{Protocol via Threshold FHE:\cite{AsharovJLTVW12}} This is a protocol that allows multiparty computation using FHE. It consists of a key generation protocol where all the parties involved collaboratively agree on a common public key of an FHE scheme and each party also receives a share of the secret key. The parties can then encrypt their individual inputs under the common public key, evaluate the desired function homomorphically on the ciphertexts, and collaboratively execute a decryption protocol on the result to learn the output of the computation. We can categorize the scheme to consist of:
\begin{itemize}
\item Threshold Key Generation.
\item Encryption and Evaluation.
\item Threshold decryption.
\end{itemize}
It is possible to convert any FHE scheme into TFHE by efficient implementation of the key-generation and decryption protocols.

We also define a variant via Threshold FHE using {\em Multi-key FHE} (MFHE) \cite{MW16} which can be used for our protocols as follows:
\begin{enumerate}
\item Each party individually chooses its own MFHE key pair $(pk_i, sk_i)$, encrypts its input $x_i$ under $pk_i$, and broadcasts the resulting ciphertext. At the end of this round, each party can homomorphically compute the desired function $f$ on the received ciphertexts and derive a common multi-key ciphertext which encrypts the output $y = f(x_1, . . . , x_n)$. Note that no individual party or incomplete set of parties can decrypt this ciphertext and so the privacy of the input is maintained.
\item The parties run a secure distributed protocol for "threshold decryption" using their secret keys $sk_i$ to decrypt the multi-key ciphertext and recover the output $y$ in plaintext.
\end{enumerate}
\noindent \textbf{Zero Knowledge Proofs:} Zero knowledge proof (refer  \cite{BFM88,GMR89,SMP87} for literature) is a technique by which a party (prover) can prove the knowledge of a statement $x$ to another party (verifier) without revealing any other information apart from the fact that she knows $x$. Any zero-knowledge proof must satisfy three properties. We informally define them as follows:
\begin{itemize}
\item Completeness: If the statement $x$ is true, then by this property, an honest verifier will be convinced of the fact that the statement is true by an honest prover.
\item Soundness: If the statement $x$ is not true, then a cheating prover cannot convince an honest verifier that $x$ is true except with negligible probability.
\item Zero knowledge: Nothing but the truth of the statement is learnt by the verifier.
\end{itemize}
Zero knowledge proof systems can be interactive or non-interactive in nature. Non-interactive zero-knowledge is a variant of zero-knowledge proof where no interaction is necessary between a prover and a verifier. To accomplish the goal, both the prover and the verifier agree on a common reference string. A common reference string is an assumption that it is a trusted setup taken from some distribution $\mathcal{}$ that is commonly shared by all parties involved in the protocol.




We consider our public encryption scheme to be CPA-secure (security against chosen-plaintext attack (CPA)). 
For more details on security of public key encryption schemes, the reader is referred to \cite{DBLP:books/crc/KatzLindell2014}.
We consider the protocol with threshold FHE scheme of \cite{AsharovJLTVW12} to explain the public perception protocol. It can be instantiated with Learning With Errors (LWE) assumption.

\section{Monitoring Protocol}
Dishonesty needs to detected by a monitor when a buyer being observed does not behave as expected from her feedback history and the ratings. A monitor knows the rule deployed by buyers for selecting sellers and has access to the public perception vector $Q^t$. If the buyer's ratings are honest, then they are the same as her personal history. Hence, if the feedback is honest, the monitor can correctly predict which seller the buyer should choose to buy from in the following round. If this prediction proves to be incorrect, it would indicate that the buyer has been dishonest. The detailed protocol is as follows:
\begin{enumerate}
\item Each monitor entity $P_i$ involved in the system has public key $pk_i$ and secret key $sk_i$ associated with it.
\item We consider a distinguished set of monitors who are honest and constantly monitor the behavior of the buyers for any corrupt behavior. Each monitor is associated with one buyer.
\item To protect the anonymity of the buyers, the SE chooses an alternate ID for each buyer and sends a copy of it to the corresponding buyer. Buyers use their alternate ID for simplified communication. The SE acts as an intermediary between the buyer and her corresponding monitor, i.e., the SE has knowledge of the association between the actual ID and alternate ID of each buyer. The monitor obtains the alternate ID of the buyer that it is monitoring and has no information about the underlying buyer. 
\item \textbf{Pseudonymity of Parties:}We allow the Software Entity to sample a random ID for each buyer involved in the computation. This is to enable ease of communication. The SE directs each buyer to include the random ID sampled as part of the message sent to the monitor i.e., both the random ID and the encrypted feedback are together sent to each monitor by the corresponding buyer. This allows the monitor to use the ID to report to the SE about any dishonest behaviour of the buyer.
\item Every feedback given by the buyer $b$ is encrypted and sent using the public key of its corresponding monitor $m$ as directed by the SE. The resulting ciphertext is decrypted by the monitor $m$ using its secret key to obtain the feedback in decrypted form for monitoring.

\item A monitor constantly observes its buyer's feedback after every purchase. If the observed behavior does not match with the feedback given by the buyer and the public perception, the monitor informs the SE about the buyer using the alternate ID and the SE publicly identifies the corrupt buyer using the actual ID of the buyer as it knows the association between the IDs.
\end{enumerate}

\section{Public Perception Protocol}
The purpose of the monitoring protocol is to ensure that the public perception vector computed is reliable. We now describe how this vector is computed, while preserving the anonymity of buyer feedback. This protocol is run amongst all the buyers on the platform. The SE also chooses a random value for $\delta _M$ necessary in each round $i$ needed for public perception vector operations. We assume that the value $\delta_M$ and trusted setup (common reference string) necessary for key generation in threshold FHE is generated and distributed by the software entity.
\begin{enumerate}
\item All buyers involved in the computation run the threshold \textbf{Key Generation} protocol of \textit{threshold fully homomorphic encryption} (\cite{AsharovJLTVW12}) to obtain their corresponding public key and private key ($pk_i, sk_i$) pair. 
\item The public keys of all the buyers are consolidated as part of the key generation protocol to obtain the common public key $pk^*$ that will be used to perform homomorphic encryption and operation on the encrypted data homomorphically. 
\item Each buyer $b_j$, $j \in [n_B]$ consolidates her feedback for the sellers using a vector $v_{b_j}$=($f_{s_1}, \cdots , f_{s_{n_S}}$) where $f_{s_1}, \cdots , f_{s_{n_S}}$ indicate the feedback values corresponding to all $n_S$ sellers $s_1$, ..., $s_n$ respectively. The feedback is -1 for low and 1 for high if $b_j$ made a purchase from $s_j$. Otherwise, $f_{s_j}=0$.
\item Each buyer $b_j$ encrypts her feedback vector $v_{b_j}$ using the consolidated public key $pk^*$ to obtain the ciphertext $c_{b_j}$ and sends this encrypted feedback to all the other buyers. $b_j$, $j \in [n_B]$ also provides a zero knowledge proof (  \cite{BFM88,GMR89,SMP87}) that the feedback input provided to compute public perception vector is the same as the one provided for monitoring purposes. The rest of the buyers as well as the SE work together as validators to ensure that the two vectors sent are the same.
\item Buyers homomorphically consolidate the vectored ciphertexts $c_{b_j}$ as rows in order to compute the matrix $F$. 

\item To separate the positive and negative feedback, the following operations are performed:
\[F_{high} = \frac{1}{2} (F + |F|)\]
\[F_{low} = \frac{1}{2} (|F| - F)\]
\item Here $|F|$ denotes the matrix $F$ with absolute values of the entries. The resulting matrices $F_{low}$ and $F_{high}$ contain 1-entries (in encrypted form) at positions where the feedback is low and high respectively. 
\item Obtaining each column sum of $F_{low}$ would give the sum of all sales made by corresponding seller with feedback of $-1$ denoted as ${sum^p_{low}}$, $p \in [n_S]$ and each column sum of $F_{high}$ would give the sum of all sales made by the corresponding seller with feedback of $1$ denoted as ${sum^p_{high}}$, $p \in [n_S]$. Total sales made by each seller $s_p$ is obtained as:
\[sum_{s_p} = {sum^p_{low}} + {sum^p_{high}}\]
Thus, $I_Q(s,i)$ = ${sum^p_{high}}/sum_{s_p}$ is obtained in encrypted form.
\item Buyers together perform distributed threshold homomorphic decryption to obtain the $I_Q$ vector for round $i$. \

\item Each buyer locally computes the public perception vector $Q^t$, using $\delta_M$ obtained from SE and verifies with other buyers that the resulting values are consistent, through consensus on the blockchain. SE chooses $\delta _M$ randomly in each round, from $(\bar{\tau} ,1)$.
\end{enumerate}
\noindent This protocol may be instantiated using the threshold multi-key fully homomorphic encryption schemes that utilize separate public key and secret keys held by the parties as well. Besides the scheme described above, we could adopt an alternative approach of using a smart contract amongst the buyers in the market, where each buyer submits their encrypted feedback, receives the $I_Q(s,i)$ vector in encrypted form and then decrypts it as in step 6 of the protocol. 

\section{Properties of the Protocols}
The regulation protocols discussed above ensure the anonymity constraints are met in all cases except a few corner cases.

\subsection*{Anonymity of Buying Patterns} 

The monitoring protocol does not reveal buying patterns. Each monitor is mapped to exactly one buyer whose identity is concealed from the monitor. As a result, the monitor can only decipher that a particular seller made a sale in that round, which is always publicly known. Feedback vectors are always sent in an encrypted format. It is ensured that whenever this encrypted vector is received from a known party, the vector cannot be decrypted by a single party. 

The cases in which a buying pattern may become public is the case when exactly one or two sellers have made a sale in a given round. In the case that there are exactly two sellers who have made any sales, {\em only these sellers} are able to decipher who bought from whom. These two corner cases are unfortunately unavoidable in our framework.

\subsection*{Anonymity of Feedback}
Despite monitors having access to the feedback vectors, anonymity is preserved, as the monitors remain unaware of the identity of the buyers whose feedback they monitor.
The feedback used in the public perception protocol is encrypted and aggregated homomorphically. As a result, buyers have no access to the feedback vector of any other buyer. The only case in which the feedback may be revealed to the seller is when his rating is either 0 or 1. This would be a consequence of all buyers giving the same feedback. These two corner cases, unfortunately  are unavoidable with in our setup. The only way to circumvent this situation would be to introduce a random offset in the ratings, which would make the system less reliable and introduce non-deterministic behavior.

\section{Conclusions}
Using homomorphic encryption and zero-knowledge proof as primitives, a regulation protocol composed of the monitoring protocol and the public perception protocol was defined. The regulation protocol provides the final building block required to facilitate a trusted reputation system amongst enterprise agents. The monitoring protocol ensures that buyers are honest in the feedback they provide and thus, the ratings computed are reliable. The public perception protocol computes the ratings while preserving the anonymity of the feedback. The protocols presented in this chapter aim to show the feasibility of achieving the requirements of a B2B platform, but do not claim to be the optimal way to do so. The enterprise procurement process, is a long one, where purchases are made every few months. Thus, inefficiency of time can be well tolerated in such systems. 
\blankpage

\chapter{Summary and Future Work}

\section{Summary }
This thesis presented a novel foray into the study of the intersection of game theory and blockchains. While blockchains are far too recent, for anything to be truly `traditional', the traditional view of the relevance of game theory to blockchains was with regards to mining and cryptoeconomic incentives. To the best of our knowledge, the work presented in this thesis is the first to use game theory to design smart contracts. 

This thesis first presented an overview of blockchains and blockchain technology in Chapter 2. Starting with the building block of blockchains, covering the data structure, organization and storage of the contents of the blockchain, along with a brief overview of mining and some of its commonly followed techniques. Key features of blockchain technology were discusses, how the blockchain functions as an asset ledger, along with the basics of smart contracts. To conclude the overview of blockchains, a common classification of blockchains into permissionless (public) and permissioned (private) blockchains, was discussed, along with the various issues associated with both.

\subsubsection*{Contributions}
In this thesis, we designed a blockchain-based market platform, which allows for B2B collaboration, discussed in Chapter 3. This platform is intended for the enterprise procurement process. Enterprises would like to choose buy their required goods from the seller, who is most likely to give them high-quality. In order to do so, they need both their own personal experience, along with reviews of other enterprise buyers. To serve this purpose, the market platform allows buyers to buy from the seller of their choice and consequently give feedback on the quality they received, so as to rate the corresponding seller. 

The collection and aggregation of this feedback also has constraints. As businesses cannot afford have their buying history leaked, for fear of revealing business strategy to rivals, the feedback must be collected such that the identities of the buyers are not revealed. Further, it is necessary to ensure that sellers are not decipher the exact feedback given by a specific buyer in a particular round. This level of anonymity can induce a buyer to be strategically dishonest about her feedback regarding a certain seller, so as to induce that seller to give high-quality, in an effort to increase his ratings, or reduce his selling price. Thus, an additional requirement, along with complete anonymity of the feedback is that buyers are incentivized to be honest.

\subsubsection*{Regulation Protocol}
These requirements were met through cryptographic protocols deployed in the form of blockchains on the platform by way of smart contracts. These protocols are detailed in Chapter 5. There are essentially two sub-protocols, the monitoring protocol and the public perception protocol, that together form the 'regulation protocol' which make the platform usable for B2B collaboration. 

In the public perception protocol, buyers on the platform, send their feedback vectors in an encrypted format to all the other buyers. The encryption scheme is homomorphic, allowing them to aggregate the data into a vector, without decrypting the contents. This vector is then decrypted and buyers locally compute the rating vector for the current round. In the monitoring protocol, each buyer's feedback is sent to a monitor, through the SE, encrypted with the monitor's public key. A monitor is a software entity itself, and does not know the identity of the corresponding buyer, identifying her only with an alternateId. The monitor uses this alternateId to alert the SE, in case the buyer is found to be dishonest.

These protocols ensure that a reliable ratings vector is computed in each round. This vector is used by the buyers to estimate which seller is most likely to give high-quality. We conduct game theoretic analysis to characterize how to best incentivize sellers to choose to produce high-quality goods. The purpose of this analysis is to help a mechanism designer ensure that the rules and parameters of the market platform are conducive to high-quality offerings in equilibrium.

\subsubsection*{Game Theoretic Analysis}
We first studied a pricing rule where all sellers charge the same price. We found that in such a pricing model, the only pure strategy subgame perfect equilibrium possible, if any, will be either giving out only low quality or only high-quality. We further deduced that {\em profit margin} and {\em competition} are necessary to induce high-quality offerings. Within homogeneous pricing, two different punishment models were studied: the ``local punishment'' model, where punishment is given locally for each infraction made by the seller, and the ``threshold punishment'' model where sellers are punished only when their ratings drop below a certain threshold. Out of the two, it was seen that the threshold punishment model is comparatively better at disincentivizing low-quality offerings. 

We then studied a case where the prices charged by sellers can be one of two prices. Within this case, we explored both types of pricing rules, where the price charged by the sellers did and did not change. For both we found that given the parameters of the system, the binary pricing case would reduce to a case of homogeneous pricing with certain parameters. We found that under any punishment rule in discrete pricing, a seller giving only high-quality, would never lose buyers, and when all other sellers give only low-quality, a seller giving high-quality, even sporadically, would be preferred by all the buyers.

It was found that these two conditions do not hold in the case of continuous pricing, where the price charged by a seller, changes even with small changes in the seller's public perception. When continuous pricing is adopted, behaviour of buyers is quite different from the behaviour observed in discrete pricing. A seller with a poorer rating may be preferred by certain buyers over a seller with a higher rating, as the former's price would be lower. As a result of this behaviour, a straightforward analysis of equilibrium behaviour is not possible for continuous pricing.

\section{Future Work }

While the work presented in this thesis aims to be as comprehensive and cover as many practical aspects as possible, there still remains scope for further work.

\subsection*{Heterogeneous Buyers}
All buyers have so far been assumed to be homogeneous. Each buyer buys the same quantity of the good in each round, and has the same value for the good. While, the case of each buyer having non-identical, but consistent demand for the good can be reduced to the model studied, with each buyer corresponding to as many pseudo-buyers, as the units of good she demands, other case cannot be captured. The case of buyers having inconsistent demand across rounds is interesting in itself.

Further, when buyers do not have the same valuation for the goods, defining a threshold for the threshold punishment model itself becomes an interesting question. Also, when buyers do not have identical valuations, it is not necessary that $v_H-p_H$ be greater than $v_L-p_L$. The analysis gets further complicated when this relation holds for a fraction of buyers and does not for others.

The assumption that all buyers are identical to sellers need not be true itself. Certain buyers may be more influential than others, and consequently, sellers may try to sustain their costs by giving high-quality to important buyers and low-quality to less important ones. Thus, buyers  being heterogeneous is an intriguing direction of future work.

\subsection*{Economies of Scale and Seller Capacities}

The model presented in this paper assumes that each seller can produce as many goods as requested, and the cost of producing each good is fixed. This is not consistent with commonly studied economic models. Microeconomic analysis commonly studies a case of {\em Economies of Scale}, where the cost of producing each additional unit of a good initially decreases and then becomes constant/increases. This case is demonstrated in Figure 6.1. This variability is not captured in the model discussed, but is essential to completely capture real-world settings.

\begin{figure}[h!]
\label{fig:eos}
\hspace{0.1\linewidth} 
\includegraphics[width=0.8\linewidth]{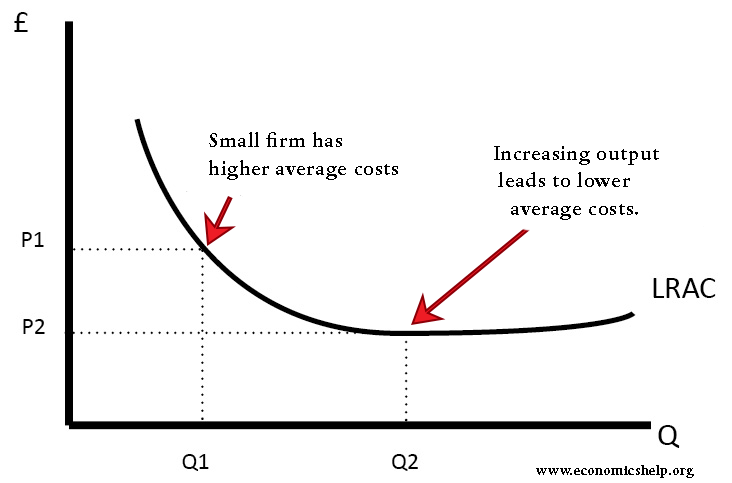}
\caption{Economies of Scale \cite{EoS}}
\end{figure}
\noindent Like costs are not fixed, neither is the additional profit from each unit of good sold. The famous ``Law of Diminishing Returns'' states that there is always a point beyond which it is no longer profitable to sell an additional unit of the good. Thus, for each seller, selling a particular good, there is a {\em capacity} beyond which it is not profitable for the seller to sell. Incorporating such capacities makes the selection of a seller by a buyer less straightforward, giving scope for changes to the model presented.

Also, the purpose of the analysis presented was to maximize the high-quality offerings, without any regards to the revenue generated. A study of a  setting which aims to achieve an optimal tradeoff between the two will be relevant.

\subsection*{Uncertainty in the Quality Produced}

Throughout the analysis we assume that a seller has complete control over the quality delivered to the buyer. The model presented does not take into consideration possibilities of error, which are beyond the control of the seller. This error may stem from manufacturing defects or from a mistake on the part of the logistics provider. However, if the model were to recognize possibility of non-strategic reasons behind a buyer receiving low-quality, the punishment models would need to be modified. 

While the threshold punishment model does not need any further restructuring to address such a case, the local punishment model may be unnecessarily harsh for this setting. Punishing every individual infraction would not be appropriate. One possible punishment method would be to punish the seller with the probability of receiving high-quality, when the seller intends to produce high-quality. Under such a punishment rule, the analysis would change drastically, even for cases that presently seem to be relatively  straightforward.

\subsection*{Differing Initial Perceptions}
The analysis of Binary Pricing rules assumes that a buyer has the same perception for all sellers, irrespective of the price they are selling at need not be true. Logically, a seller selling a good at a higher price should guarantee better quality. Thus, differing values of $\xi$ for different price brackets, with buyers further having different values of $\xi$ amongst each other as well. 

The assumption of the quality of a product being one of two states itself is somewhat harsh. Quality of a product may be one of many discrete states, or may even belong to a continuous interval. For example the duration before the product wears out is a good indication of its quality. In such a setting, valuations will also change from a set of values to a function, giving scope for further work in this direction.

\subsection*{Learning Mechanisms}
The work discussed in this thesis is purely analytical. We assume that the parameters of the system are correctly known by all the agents, and agents even know their own private values correctly. The latter need not be true in the case of sellers. A seller need not accurately know the quality that he is capable of providing, and further need not know the price that he should be charging. 

In such a case, it would be appropriate to have a learning mechanism which, through the feedback provided by the buyers, aims to learn the quality of the sellers and accordingly elicit a price corresponding to each seller. Such a mechanism can be implemented by means of a smart contract on the platform, in conjunction with the regulation protocols, so as to maintain the anonymity requirements. As a step towards achieving such a mechanism, a mechanism design formulation of the problem studied in this thesis would be required.

\subsection*{Decentralizing the Software Entity}

Currently, the software entity plays a single point of centralization in the monitoring protocol. This doesn't fit very well within the decentralized paradigm intended by blockchains. A significant improvement would be to have a distributed manner of matching buyers with monitors without leaking the identity of the buyers to the monitors. Eliminating the need of zero-knowledge proofs in this system would be another possible line of future work.

\subsection*{Implementation}
The implementation of a B2B market platform, as studied in this thesis is of independent interest. As discussed, the implementation is very much feasible on currently available blockchain technology, such as Hyperledger Fabric. It is, however, non-trivial. While, private channels between a buyer and seller and well-established on such platforms, providing for inter-operable smart contracts is not easy to accomplish. Further, implementing MPC protocols on a blockchain-based framework are a novel contribution in itself. The implementation using permissioned blockchains to deploy smart contracts to compute reliable ratings and incentivize sellers to provide high-quality is necessary to bring to completion the work presented in this thesis. This we believe is an important direction for translating this work into a real-world system.




\bibliographystyle{plain}
\bibliography{references}
\end{document}